\documentclass[article,amsmath,amssymb]{revtex4-2}
\usepackage{graphicx}
\usepackage{dcolumn}
\usepackage[colorlinks]{hyperref}
\usepackage{bm}

\begin{document}
\title{Subradiance generation in a chain of two-level atoms with a single excitation}

\author{Nicola Piovella}
\affiliation{Dipartimento di Fisica "Aldo Pontremoli", Universit\`{a} degli Studi di Milano, Via Celoria 16, I-20133 Milano, Italy \&
INFN Sezione di Milano, Via Celoria 16, I-20133 Milano, Italy}

\begin{abstract}
Studies of subradiance in a chain $N$  two-level atoms in the single excitation regime focused mainly on the complex spectrum of the effective Hamiltonian, identifying subradiant eigenvalues. This can be achieved by finding the eigenvalues $N$ of the Hamiltonian or by evaluating the expectation value of the Hamiltonian on a generalized Dicke state, depending on a continuous variable $k$. This has the advantage that the sum above $N$ can be calculated exactly, such that $N$ becomes a simple parameter of the system and no more the size of the Hilbert space. However, the question remains how subradiance emerges from atoms initially excited or driven by a laser. Here we study the dynamics of the system, solving the coupled-dipole equations for $N$ atoms and evaluating the probability to be in a generalized Dicke state at a given time. Once the subradiant regions has been identified, it is simple to see if subradiance is being generated. We discuss different initial excitation conditions that lead to subradiance and the case of atoms excited by switching on and off a weak laser. This may be relevant for future experiments aimed at detecting subradiance in ordered systems.
\end{abstract}

\maketitle

\section{Introduction}

Since the seminal paper by Dicke \cite{Dicke1954}, collective spontaneous emission  by an ensemble of two-level atoms has been studied by many authors \cite{Lehmberg1970,Bonifacio1971,Gross1982}. In particular subradiance, i.e. inhibited emission due to destructive interference between the emitters, has recently received great attention, both in disordered systems as in a cloud \cite{Bienaime2012,Guerin2016,Das2020,Ferioli2021}, and in ordered systems, as in atomic chains or 2D and 3D lattices \cite{Bettles2016,Facchinetti2016,Asenjo2017,Rui2020,Cech2023,Piovella2024}. The theoretical studies on subradiance has been focused mostly in the study to the eigenvalues of the system \cite{Bellando2014,Cottier2018} and in particular on the collective decay rate, which for subradiance is less than the single-atom decay $\Gamma$. 

In disordered system, subradiance has been investigated numerically and experimentally by switching off a  continuous detuned laser driving a thermal cloud and calculating the decay rate of the fluorescence light intensity by reporting it in a semilog plot \cite{Bienaime2012,Guerin2016,Fofanov2021}.
After the initial fast decay, subradiance manifests itself in a slowly decaying fluorescence intensity with a rate below the single atom decay rate. 
As a consequence of the existence of a single superradiant mode and many degenerate subradiant modes, 
at first the subradiant decay is not purely exponential, since several modes decay simultaneously. For longer times, it then ends up with a pure exponential decay (referred as subradiant decay) when only one long-lived mode dominates. A similar approach can be adopted for ordered system, as atoms in a linear chain. Such linear chains have been investigated theoretically by different authors \cite{Nienhuis1987,Zoubi2010,Jenkins2012,Bettles2015,Needham2019,Masson2020,Masson2022,Ruostekoski2023}. In the single-excitation approximation, subradiance has been studied in the previous literature by calculating the eigenvalues of the system, determined numerically by diagonalizing the finite $N\times N$ matrix associated with the Green operator describing the coupling between the emitters.

Recently, we proposed a different method to study subradiance in a finite linear chain, based on the evaluation of the expectation value of the effective Hamiltonian on a generalized Dicke state, depending on a continuous variable $k$ \cite{Piovella2024}. This has several advantages: 1) it allows to calculate the collective decay rate $\Gamma(k)$ as a function of a continuous parameter, mimic the exact Fourier spectrum; 2) The sum over $N$ can be calculated exactly, making $N$ a mere parameter of the system and no more the dimension of the matrix whose eigenvalues are to be evaluated; 3) the study of $\Gamma(k)$ allows us to identify the subradiant regions of the spectrum. The last point is important, because it can be used to detect how subradiance may be dynamically generated. 

In general, little attention has been devoted to the generation of subradiance, from the preparation of the excited atoms or when the atoms are driven by a laser. Few exceptions are provided for instance by ref.\cite{Rui2020}, where the cooperative subradiant response of a two-dimensional square array of atoms in an optical lattice has been observed. Other experimental demonstration of subradiance have been reported in \cite{Jenkins2017,Solano2017,Ferioli2021}. A detailed description of methods commonly
employed to analyze the cooperative responses of atomic arrays and explore some recent developments and
potential future applications of planar arrays is contained in ref.\cite{Ruostekoski2023}. The long subradiant lifetimes may be used for storage and retrieval of quantum information \cite{Facchinetti2016,Jen2016} and other photonic
devices, for instance, nanolasers \cite{Holzinger2020}, plasmonic ring nanocavities\cite{Sonnefraud2010} and ultracold molecules \cite{McGuyer2015}.

By solving the coupled-dipole equations for $N$ atoms with given initial conditions or in presence of a driving laser, it is possible to project the solution at a given time on the generalized Dicke state. The result gives the probability distribution of the state as a function of the continuous variable $k$. Then, by identifying the subradiant regions of the spectrum, it is possible to see if subradiance has been generated. In particular, we find which is the initial state maximally generating subradiance: this is useful in order to understand the symmetry properties of the subradiant state.

We assume here an ideal chain. In a real experiment fluctuations in the atomic position may be detrimental for cooperative effects \cite{Morsch2006, Hadzibabic2004}. For instance, the role of imperfections of the collective decay in a 1D array has been studied in ref.\cite{Masson2020}, showing that these phenomena are robust to realistic experimental imperfections.

The paper is organized as follow. In the first part we review the main results of Ref.\cite{Piovella2024}, defining the generalized Dicke state and adding the calculation of the collective frequency shift, both in the scalar and vectorial models.  The second part is devoted to the generation of subradiance by suitable initial atomic excitations or when atoms are excited by a weak laser field. The dynamics of the system is investigated  by solving the $N$ coupled-dipole equations and projecting the single-particle state a the time $t$ over the generalized Dicke state. We will see that it is possible to build a maximally subradiant state, which in the limit of an infinite chain gives a vanishing spontaneous decay rate. Finally, we interpret the results in terms of the fluorescence intensity spatial distribution.

\section{Modelling emission from a chain of atom with a single excitation}

Here we define the   collective frequency shift and decay rate as the real and imaginary part of the expectation value of the effective Hamiltonian over the generalized Dicke state. Preliminary results have been previously reported in Ref. \cite{Piovella2024}. The calculations are carried on first for the scalar model and then extended to the vectorial model in sec.\ref{vec:model}.

\subsection{Scalar model}

We consider $N$ two-level atoms  with ground state $|g_j\rangle$ and excited state $|e_j\rangle$ ($j=1,\ldots,N$), with atomic transition frequency $\omega_0=ck_0$, linewidth $\Gamma$, dipole $\mu$ and position $\mathbf{r}_j$.
We consider here the single-excitation effective Hamiltonian in the scalar approximation \cite{Akkermans2008,Bienaime2013}
\begin{eqnarray}
\hat H & =&-i\frac{\hbar}{2}\sum_{j,m}G_{jm}\,
    \hat{\sigma}_j^\dagger\hat\sigma_m,\label{Heff}
\end{eqnarray}
 where $\hat\sigma_j=|g_j\rangle\langle e_j|$ and $\hat\sigma_j^\dagger=|e_j\rangle\langle g_j|$ are the lowering and raising operators, $G_{jm}$ is the scalar Green function,
\begin{equation}\label{gammajm}
    G_{jm}=
   	\left\{
    \begin{array}{ll}
    \Gamma_{jm}-i\, \Omega_{jm} & \mbox{if}~j\neq m, \\[1ex]
    \Gamma & \mbox{if}~j = m,
\end{array}
	\right.
\end{equation}
and
\begin{equation}\label{gammajm:bis}
    \Gamma_{jm} = \Gamma\frac{\sin(k_0r_{jm})}{k_0r_{jm}}\quad , \quad
    \Omega_{jm} = \Gamma\frac{\cos(k_0r_{jm})}{k_0r_{jm}},
\end{equation}
where $r_{jm}=|\mathbf{r}_j-\mathbf{r}_m|$. $\Gamma_{jm}$ can be obtained as the angular average of the radiation field propagating between the  two atomic positions $\mathbf{r}_j$ and $\mathbf{r}_m$ with wave-vector $\mathbf{k}=k_0(\sin\theta\cos\phi,\sin\theta\sin\phi,\cos\theta)$ (see Appendix \ref{eq:4}),
\begin{equation}
\Gamma_{jm}=\frac{\Gamma}{2}\left\langle e^{-i\mathbf{k}\cdot (\mathbf{r}_j-\mathbf{r}_m)} 
+\mathrm{c.c.}\right\rangle_\Omega\label{Sk}
\end{equation}
where the angular average is defined as
\[
\left\langle f(\theta,\phi)\right\rangle_\Omega=\frac{1}{4\pi}\int_0^{2\pi}d\phi\int_0^\pi \sin\theta f(\theta,\phi)d\theta.
\]
Eq.(\ref{Sk}) provides a simple interpretation of $\Gamma_{jm}$ as the coupling between the $j$th atom and the $m$th atom, mediated by the photon shared between the two atoms and averaged over all the vacuum modes. Eq.(\ref{Sk}) allows to factorize $\Gamma_{jm}$ in the product of two terms, before averaging them over the total solid angle.

We consider $N$ atoms placed along a linear chain with lattice constant  $d$, with $\mathbf{r}_j=d(j-1)\mathbf{\hat e}_z$, with $j=1,\dots ,N$, so that Eq.(\ref{Sk}) becomes
\begin{equation}
\Gamma_{jm}=\frac{\Gamma}{4}\int_0^\pi\sin\theta\left[e^{-ik_0d(j-m)\cos\theta}+\mathrm{c.c.}\right]d\theta.\label{Gamma:1D}
\end{equation}

\subsection{Generalized Dicke state.}

We define the generalized Dicke states \cite{Piovella2024}
 \begin{equation}
 |k\rangle =\frac{1}{\sqrt{N}}\sum_{j=1}^N e^{ikd(j-1)}|j\rangle\label{Dicke}
 \end{equation}
 where $|j\rangle=|g_1,\dots, e_j,\dots, g_N\rangle$ and $k\in (-\pi/d, \pi/d)$. It includes for $k=0$ the Dicke state \cite{Dicke1954} and for $k=k_0$ the Timed-Dicke state introduced by Scully \cite{Scully2006,Scully2015}. It satisfies the completeness relation
 \begin{equation}
\frac{d}{2\pi}\int_{-\pi/d}^{\pi/d}dk |k\rangle\langle k|=1.
 \end{equation}
As expected, the states $|k\rangle$ are not orthogonal for a finite chain, since
\begin{equation}
\langle k'|k\rangle=\frac{\sin[(k-k')dN/2]}{\sin[(k-k')d/2]}e^{i(k-k')d(N-1)/2},
\end{equation}
but they become so for an infinite chain, $\langle k'|k\rangle\rightarrow \delta(k-k')$ for $N\rightarrow\infty$. So the states $|k\rangle$ form an over-completed basis for the single-excitation manifold.

Notice that it is possible to extend the definition of the generalize Dicke state (\ref{Dicke}) to the multi-excitation manifolds. Then, the approach described in the next sections can be straightforwardly extended to the multi-excitations modes. For instance, a two-excitation generalized Dicke state may be defined as
\begin{equation}
 |k_1,k_2\rangle ={\cal C}\sum_{j=1}^N\sum_{m=1\atop m\neq j}^N e^{ik_1d(j-1)+ik_2d(m-1)}|j,m\rangle\label{two-Dicke}
\end{equation}
where $|j,m\rangle=|g_1,\dots, e_j,\dots,e_m,\dots, g_N\rangle$ and ${\cal C}$ is a normalization constant. Preliminary results about two-excitation modes have been discussed in ref.\cite{Asenjo2017}.

\subsection{Collective frequency shift and decay rate.}

Taking the expectation value of the effective Hamiltonian (\ref{Heff}) over the generalized Dicke state (\ref{Dicke}) yields
\begin{equation}
-\frac{2}{\hbar}\langle k|\hat H|k\rangle=\Omega_N(k)+i\Gamma_N(k)
\end{equation}
were 
\begin{equation}
\Omega_N(k)=\frac{1}{N}\sum_{j=1}^N\sum_{m=1\atop m\neq j}^N\Omega_{jm}e^{ikd(j-m)}\label{Omega:k}
\end{equation}
is the collective frequency shift and 
\begin{equation}
\Gamma_N(k)=\frac{1}{N}\sum_{j=1}^N\sum_{m=1}^N\Gamma_{jm}e^{ikd(j-m)}\label{Gamma:k}
\end{equation}
is the collective decay rate. By using Eq.(\ref{Gamma:1D}) in Eq.(\ref{Gamma:k}) we can write
\begin{eqnarray}
  \Gamma_N(k) &=&\frac{\Gamma}{2N}\int_0^\pi\sin\theta|F_k(\theta)|^2d\theta
\end{eqnarray}
where
\begin{eqnarray}
|F_k(\theta)|^2&=&\left|\sum_{j=1}^N e^{i(k-k_0\cos\theta)d(j-1)}\right|^2=\frac{\sin^2[(k-k_0\cos\theta)dN/2]}{\sin^2[(k-k_0\cos\theta)d/2]}
\label{Fk}
\end{eqnarray}
and
\begin{eqnarray}
  \Gamma_N(k) &=&\frac{\Gamma}{k_0dN}\int_{(k-k_0)d/2}^{(k+k_0)d/2}\frac{\sin^2(Nt)}{\sin^2 t}dt\label{Gk:1}
\end{eqnarray}
where we changed the integration variable from $\theta$ to $t=(k-k_0\cos\theta)d/2$.
For large $N$, we can approximate in the integral of Eq.(\ref{Gk:1}),
\begin{equation}
\frac{\sin^2(Nt)}{\sin^2t}\approx N^2\sum_{m=-\infty}^{+\infty} \mathrm{sinc}^2\left[\left(t-m\pi\right)N\right],\label{approx}
\end{equation}
where $\mathrm{sinc}(x)=\sin x/x$,
so that
\begin{eqnarray}
  \Gamma_N(x) &=&
  =\frac{\Gamma N}{a}\sum_{m=-\infty}^{+\infty} \int_{(x-a)/2}^{(x+a)/2}\mathrm{sinc}^2\left[\left(t-m\pi\right)N\right]dt.\label{Gamma:scalar}
\end{eqnarray}
where $a=k_0d$ and $x=kd$, with $x\in(-\pi,\pi)$. Hence, we have transformed the double sum in Eq.(\ref{Gamma:k}) into an integral, where $N$ plays the role of a simple parameter of the system.
The collective frequency shift takes the form
\begin{equation}
\Omega_N(x)=\frac{\Gamma}{N}\sum_{j=1}^N\sum_{m=1\atop m\neq j}^N\frac{\cos(a[j-m|)}{a|j-m|}e^{ix(j-m)}
\label{Omega:scalar}.
\end{equation}

\subsection{Infinite chain}

If the chain is infinite, $N\rightarrow\infty$, the solution for the collective decay rate is
\begin{eqnarray}
  \Gamma_{\infty}(x) &=&\frac{\Gamma\pi}{a}\sum_{m=-\infty}^{+\infty} \Pi[2m\pi-a<x<2m\pi+a]\label{approx:scalar:BZ}
\end{eqnarray}
where $\Pi(a<x<b)$ is the rectangular function, equal to 1 for $a<x<b$ and 0 elsewhere. In the first Brillouin zone, $m=0$, $\Gamma_{\infty}(x)=\Gamma\pi/a$ for $|x|<a$ and $\Gamma_{\infty}(x)=0$ for $a<|x|<\pi$. Atomic modes in the region enclosed
within the light line $k=\pm k_0$ are generally unguided and radiate into free space. Outside the light line ($|k|>k_0$), the modes are
guided and subradiant, as the electromagnetic field is evanescent in the directions transverse to the chain.
For an infinite chain, $\lim_{N\rightarrow\infty}\Omega_N(x)=\Omega_\infty(x)$ depends only on the index $\ell=j-m$,
\begin{eqnarray}
\Omega_\infty(x)&=&\Gamma\sum_{\ell=-\infty}^\infty\frac{\cos(a|\ell|)}{a|\ell|}e^{ix\ell}
= \frac{\Gamma}{2a}\sum_{\ell=1}^\infty\frac{1}{\ell}\left[e^{i(a+x)\ell}+e^{i(a-x)\ell}+\mathrm{c.c.}\right].
\end{eqnarray}
Using the expansion
\[
\ln(1-z)=-\sum_{n=1}^\infty
\frac{z^n}{n}
\]
we write
\begin{eqnarray}
\Omega_\infty(x)
&=&-\frac{\Gamma}{2a}\left\{\ln[1-e^{i(a+x)}]+\ln[1-e^{i(a-x)}]+\mathrm{c.c.}\right\}\nonumber\\
&=&-\frac{\Gamma}{a}\ln\left[2|\cos a -\cos x|\right]\label{WW}
\end{eqnarray}
The frequency shift has a logarithmic divergence for $x=\pm a$ and three extremes at $x=0,\pm\pi$, with $\Omega_\infty(0)=-(2\Gamma/a)\ln[2|\sin(a/2)|]$ and $\Omega_\infty(\pm\pi)=-(2\Gamma/a)\ln[2|\cos(a/2)|]$, respectively.

\subsection{Finite chain}

If the chain is finite, the collective decay rate $\Gamma_N(x)$ can be calculated by using Eq.(\ref{Gamma:scalar}). The collective frequency shift of Eq.(\ref{Omega:scalar}) can be written transforming the double sum in a single sum on the index $\ell=j-m$, with a degenerate factor $g_\ell=N-\ell$:
\begin{eqnarray}
\Omega_N(x)&=&\frac{2\Gamma}{N}\sum_{\ell=1}^{N-1}(N-\ell)\frac{\cos(a\ell)}{a\ell}\cos(x\ell)
\label{Omega:ell}
\end{eqnarray}

\subsection{Vectorial model}\label{vec:model}
 
\begin{figure}
      \centerline{\scalebox{0.4}{\includegraphics{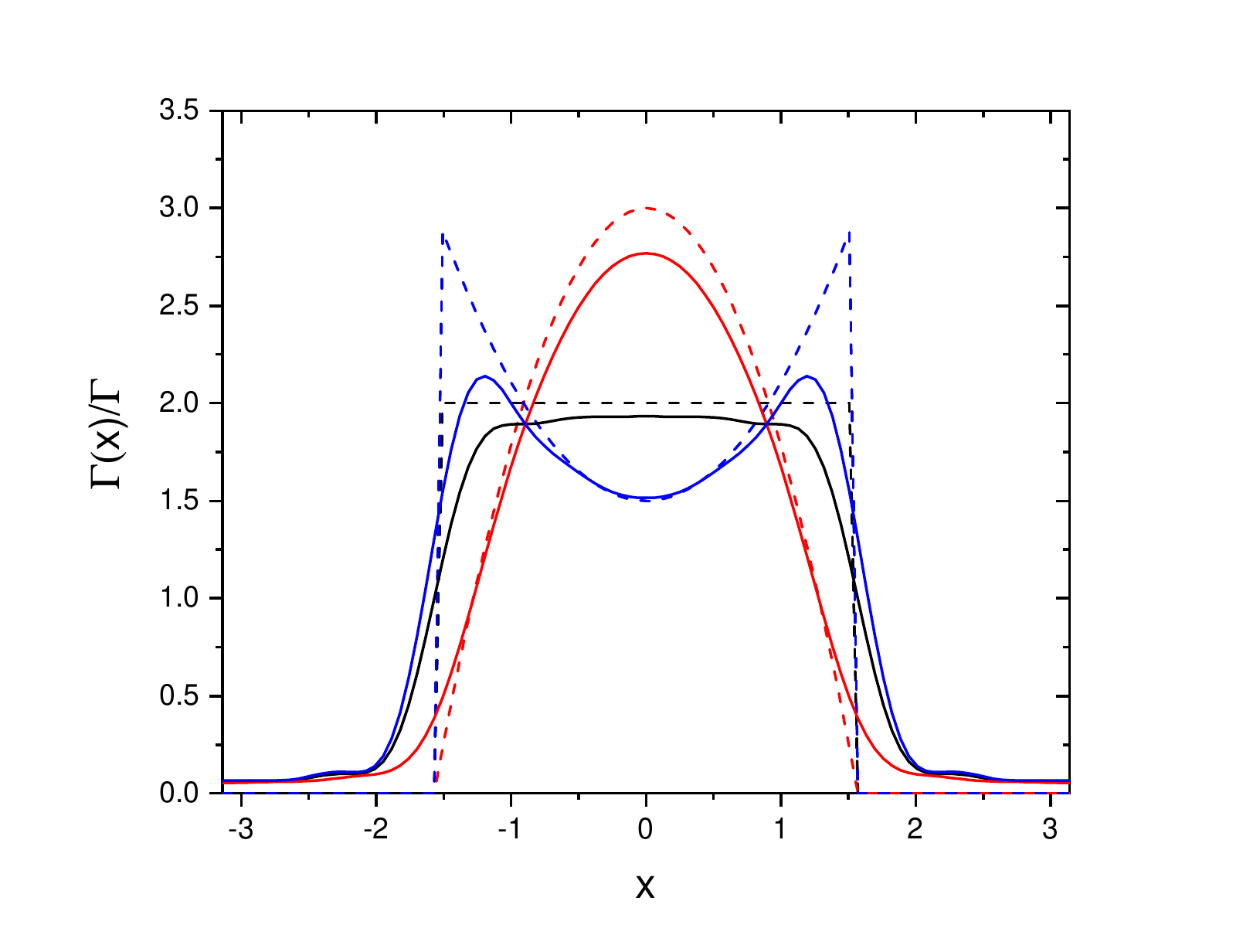}}}
        \caption{$\Gamma(x)/\Gamma$ vs $x$ for $a=\pi/2$. Dashed lines are for an infinite chain, continuous lines for a finite chain with $N=10$. Black lines are for the scalar model, red and blue lines for the vectorial model with $\delta=0$ and $\delta=\pi/2$, respectively.}
        \label{fig1}
\end{figure}

\begin{figure}
      \centerline{\scalebox{0.4}{\includegraphics{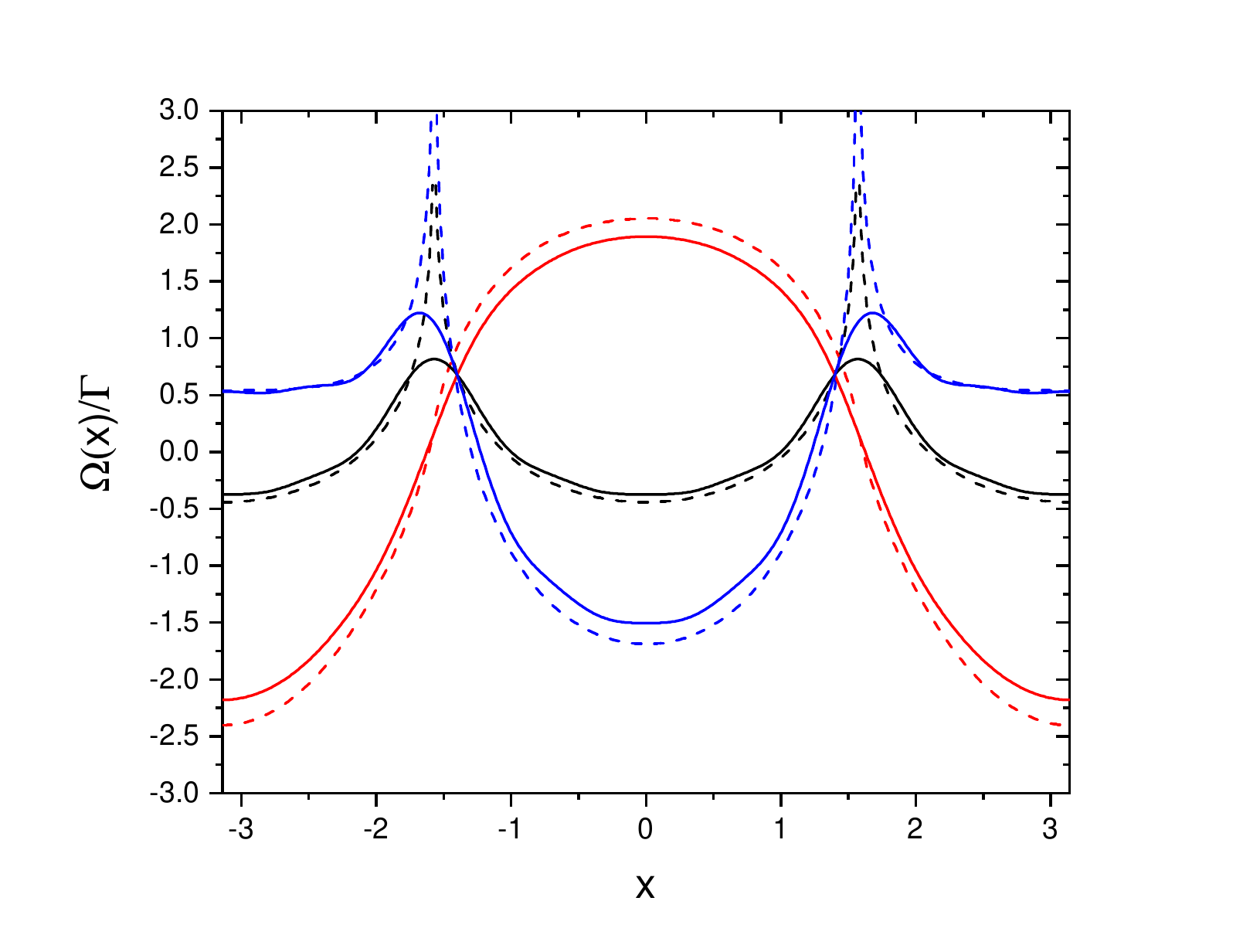}}}
        \caption{$\Omega(x)/\Gamma$ vs $x$ for $a=\pi/2$. Dashed lines are for an infinite chain, continuous lines for a finite chain with $N=10$. Black lines are for the scalar model, red and blue lines for the vectorial model with $\delta=0$ and $\delta=\pi/2$, respectively.}
        \label{fig2}
\end{figure}

We now extend the previous expressions to the vectorial model, taking into account the polarization of the electromagnetic field. The non-Hermitian Hamiltonian is now
\begin{eqnarray}
\hat H & =&-i\frac{\hbar}{2}\sum_{\alpha,\beta}\sum_{j,j'}G_{\alpha,\beta}(\mathbf{r}_j-\mathbf{r}_{j'})\,
    \hat{\sigma}_{j,\alpha}^\dagger\hat\sigma_{j',\beta}.\label{H_vec}
\end{eqnarray}
where $\alpha,\beta=(x,y,z)$. Here  $\hat\sigma_{j,x}=(\hat\sigma_{j}^{m_J=1}+\hat\sigma_{j}^{m_J=-1})/2$, $\hat\sigma_{j,y}=(\hat\sigma_{j}^{m_J=1}-\hat\sigma_{j}^{m_J=-1})/2i$ and $\hat\sigma_{j,z}=\hat\sigma_{j}^{m_J=0}$, where 
 $\hat\sigma_{j}^{m_J}=|g_j\rangle\langle e_{j}^{m_J}|$ is the lowering operator between the ground state $|g_j\rangle$ and the three excited states $|e_{j}^{m_J}\rangle$ of the $j$th atom with quantum numbers $J=1$ and $m_J=(-1,0,1)$. The vectorial Green function in Eq.(\ref{H_vec}) is 
\begin{equation}
G_{\alpha,\beta}(\mathbf{r})=\frac{3\Gamma}{2}\frac{e^{ik_0r}}{ik_0r}\left[\delta_{\alpha,\beta}-\hat n_\alpha\hat n_\beta
+\left(\delta_{\alpha,\beta}-3\hat n_\alpha\hat n_\beta\right)\left(\frac{i}{k_0 r}-\frac{1}{k_0^2 r^2}\right)
\right]
\end{equation}
with $r=|\mathbf{r}|$ and ${\hat n}_\alpha$ being the components of the unit vector $\hat{\mathbf{n}}=\mathbf{r}/r$. 
We consider the linear chain with lattice constant $d$, i.e. $\mathbf{r}_j=d(j-1)\mathbf{\hat e}_z$, with $j=1,\dots ,N$, and all the dipoles aligned with an angle $\delta$ with respect to the chain's axis, so that $\hat n_\alpha=\hat n_\beta=\cos\delta$ and
\begin{equation}
G^{(\delta)}(r_{jm})=\frac{3\Gamma}{2}\frac{e^{ik_0r_{jm}}}{ik_0r_{jm}}\left[\sin^2\delta
+(1-3\cos^2\delta)\left(\frac{i}{k_0r_{jm}}-\frac{1}{k_0^2r_{jm}^2}\right)\right]\label{Gaa}
\end{equation}
where $r_{jm}=d|j-m|$.
The decay rate for the vectorial model is given by the real part of $G^{(\delta)}(r_{jm})$,
\begin{eqnarray}
\Gamma^{(\delta)}(r_{jm})&=&
\frac{3\Gamma}{2}\left[\sin^2\delta j_0(k_0r_{jm})+(3\cos^2\delta-1)\frac{j_1(k_0r_{jm})}{k_0r_{jm}}\right]\label{Gamma:vec}
\end{eqnarray}
where $j_0(x)=\sin x/x$ and $j_1(x)=\sin x/x^2-\cos x/x$ are the spherical Bessel functions of order $n=0$ and $n=1$.
The frequency shift  is given by the negative of the imaginary part of $G^{(\delta)}(r_{jm})$,
\begin{eqnarray}
\Omega^{(\delta)}(r_{jm})&=&
\frac{3\Gamma}{2}\left[\sin^2\delta \frac{\cos(k_0r_{jm})}{k_0r_{jm}}+(3\cos^2\delta-1)\left(
\frac{\sin(k_0r_{jm})}{(k_0r_{jm})^2}+\frac{\cos(k_0r_{jm})}{(k_0r_{jm})^3}
\right)\right]\label{Omega:vec}
\end{eqnarray}
We note that Eq.(\ref{Gamma:vec}) and (\ref{Omega:vec}) reduce to the expressions (\ref{gammajm:bis}) of the scalar model for $\cos^2\delta=1/3$, i.e. for $\delta=54.73^\circ$. Hence the scalar model, generally considered unrealistic for non-dilute systems, in a linear chain can be obtained for a particular orientation of the dipoles.
As done for the scalar model, we define a collective decay rate, $\Gamma_N^{(\delta)}(k)=-(2/\hbar)\mathrm{Im}\langle k|\hat H|k\rangle$, and a collective frequency shift, $\Omega_N^{(\delta)}(k)=-(2/\hbar)\mathrm{Re}\langle k|\hat H|k\rangle$, where $\hat H$ is defined in Eq.(\ref{H_vec}) and where now $|k\rangle=(1/\sqrt{N})\sum_{j=1}^N e^{ikd(j-1)}|g_1,..,e_j^{(\delta)},..,g_N\rangle$, where $|e_j^{(\delta)}\rangle$ denotes the excited atoms with the combination of the Zeeman sublevels yielding the dipoles oriented with the angle $\delta$ with respect to the chain's axis.
It is possible to demonstrate that the collective decay rate is \cite{Piovella2024}
\begin{eqnarray}
  \Gamma_N^{(\delta)}(x) &=&\frac{3\Gamma N}{2a}\sum_{m=-\infty}^{+\infty} \int_{(x-a)/2}^{(x+a)/2}
  \left[\sin^2\delta+\frac{1}{2}(1-3\cos^2\delta)\frac{(x-2t)^2-a^2}{a^2}
  \right]\nonumber\\
  &\times &\mathrm{sinc}^2\left[\left(t-m\pi\right)N\right]dt\label{gamma_vec:2}
\end{eqnarray}
(where $x=kd$) while the collective frequency shift is
\begin{eqnarray}
\Omega_N^{(\delta)}(x)&=&\frac{2}{N}\sum_{\ell=1}^{N-1}(N-\ell)\Omega^{(\delta)}(a\ell)\cos(x\ell).
\end{eqnarray}
If the chain is infinite, $N\rightarrow\infty$, 
\begin{eqnarray}
  \Gamma_\infty^{(\delta)}(x) &=&\frac{3\Gamma\pi}{2a}\sum_{m=-\infty}^{+\infty}
  \left\{\sin^2\delta+\frac{1}{2}(1-3\cos^2\delta)\frac{(x-2\pi m)^2-a^2}{a^2}
  \right\}\nonumber\\
  &\times &\Pi[2m\pi-a<x<2m\pi+a],\label{gamma_vec:3}
\end{eqnarray}
and
\begin{eqnarray}
\Omega_\infty^{(\delta)}(x)&=&2\sum_{\ell=1}^\infty\Omega^{(\delta)}(a\ell)\cos(x\ell)\nonumber\\
&=&\frac{3\Gamma}{2a^3}\mathrm{Re}\left\{
-a^2\sin^2\delta\left[\ln\left(1-e^{i(x+a)}\right)+\ln\left(1-e^{i(x-a)}\right)\right]
+(3\cos^2\delta-1)\right.\nonumber\\
&\times &\left.\left[
-ia
\mathrm{Li}_2\left(e^{i(x+a)}\right)+ia\mathrm{Li}_2\left(e^{i(x-a)}\right)
+
\mathrm{Li}_3\left(e^{i(x+a)}\right)+\mathrm{Li}_3\left(e^{i(x-a)}\right)
\right]
\right\}
\end{eqnarray}
where  $\mathrm{Li}_\nu(z)=\sum_{\ell=1}^\infty z^{\ell}/\ell^\nu$ is the PolyLog function.
Fig.\ref{fig1} and Fig.\ref{fig2} show $\Gamma(x)/\Gamma$ and $\Omega(x)/\Gamma$ vs $x$ for $a=\pi/2$, for an infinite chain (dashed lines) and for a finite chain with $N=10$ (continuous lines). Black lines are for the scalar model, red and blue lines are for the vectorial model, with $\delta=0$ and $\delta=\pi/2$, respectively. For an infinite chain the collective decay rate is zero for $|x|>a$ (i.e. for $|k|>k_0$), both for the scalar and the vectorial model. 
Notice that for $\delta=0$ the phase shift $\Omega_\infty^{(\delta=0)}(x)$ is not diverging at $x=\pm a$.

\section{Dynamics}

Having characterized the properties of the collective decay rate and frequency shift, identifying the subradiant zone $|x|>a$ where the collective decay rate is less than the single-atom decay rate $\Gamma$, we are interested now to study how subradiance can be generated by properly exciting the atoms. First we define the probability distribution of the system to be in a given generalized Dicke state $|k\rangle$, expressed in term of the single-particle basis. Then we will study the time evolution of this distribution, leading to subradiance. The following expressions are valid for both the scalar and vectorial model, so we will omit the suffix $(\delta)$ where not necessary.

\subsection{The probability density $P(k)$}

Let us assume that the state of the system is $|\Psi\rangle=\alpha |g_1\dots g_N\rangle+|\Psi'\rangle$, where
\begin{equation}
 |\Psi'\rangle=\sum_{j=1}^N \beta_j|j\rangle
 \end{equation}
 describes the state in the single-excitation manifold.  
 By projecting on the generalized Dicke basis (\ref{Dicke}),
 \begin{equation}
 |\Psi\rangle
 =\frac{d}{2\pi}\int_{-\pi/d}^{\pi/d}dk |k\rangle\langle k|\Psi'\rangle
 \end{equation}
 where
 \begin{equation}
\langle k|\Psi'\rangle=\frac{1}{\sqrt{N}}\sum_{j=1}^N e^{-ikd(j-1)}\beta_j=\frac{1}{\sqrt{N}}A_N(k).
\label{def:A}
 \end{equation}
 Hence, the probability density to be in a state $|k\rangle$ is
\begin{equation}
 P(k)=\frac{dN}{2\pi}\frac{|\langle k|\Psi'\rangle|^2}{\langle\Psi'|\Psi'\rangle}=\frac{d}{2\pi}
 \frac{\left|\sum_{j=1}^N e^{-ikd(j-1)}\beta_j(t)\right|^2}{\sum_{j=1}^N |\beta_j|^2}
 =\frac{|A_N(k)|^2}{\int_{-\pi/d}^{\pi/d}|A_N(k)|^2dk}\label{Pk}
 \end{equation}
 with
 \begin{equation}
 \int_{-\pi/d}^{\pi/d}P(k)dk=1.
 \end{equation}
 Equation (\ref{Pk}) expresses the probability density in terms of the dipole amplitudes of the single atoms, whose time evolution is described in the following section.
 
 \subsection{Dynamics of the probability density $P(k)$}
 
Let consider the time evolution of the atomic system in the presence of an external driving field in the scalar model.
In the linear regime, the probability amplitudes $\beta_j(t)$ evolve with the following coupled-dipole equations,
 \begin{equation}
 \frac{d\beta_j}{dt}=i\Delta_0\beta_j-i\frac{\Omega_0}{2}e^{ia(j-1)}-\frac{\Gamma}{2}\sum_{m=1}^NG_{jm}\beta_m
 \label{beta_dot}\end{equation}
 where $G_{jm}$ is defined in Eq.(\ref{gammajm}),
 $a=k_0d$, $\Omega_0$ and $\Delta_0$ are the Rabi frequency and the laser-atom detuning of the driving laser field. In the vectorial model and for dipoles all aligned with the same angle $\delta$ with respect to the chain's axis, the probability amplitudes $\beta_j^{(\delta)}(t)$
evolve with the equations,
 \begin{equation}
 \dot\beta_j^{(\delta)}=\left(i\Delta_0-\frac{\Gamma}{2}\right)\beta_j^{(\delta)}-i\frac{\Omega_0^{(\delta)}}{2}e^{ia(j-1)}-\frac{1}{2}\sum_{m=1\atop m\neq j}^N G^{(\delta)}(a|j-m|)\beta_m^{(\delta)}
 \label{beta_dot:vec}\end{equation}
where the driving field has the same polarization as the dipoles and $G^{(\delta)}(x)=\Gamma^{(\delta)}(x)-i\Omega^{(\delta)}(x)$, where
$\Gamma^{(\delta)}(x)$ and $\Omega^{(\delta)}(x)$ are defined in Eqs.(\ref{Gamma:vec}) and (\ref{Omega:vec}).  We notice that Eqs.(\ref{beta_dot}) and (\ref{beta_dot:vec}) have the same form, the only difference being the expression of $G$. In the following we will use the same notation for both the scalar and vectorial model.
   
  From Eq.(\ref{beta_dot}), it is possible to obtain the equation for the temporal evolution of the probability amplitude $A_N(x,t)$ defined in Eq.(\ref{def:A}) (where $x=kd$) (see Appendix \ref{eq:A}):
\begin{eqnarray}
\frac{\partial A_N(x,t)}{\partial t}&=& \left(i\Delta_0-\frac{\Gamma}{2}\right)A_N(x,t)-i\frac{\Omega_0}{2}\frac{\sin[(x-a)N/2]}{\sin[(x-a)/2]}e^{-i(x-a)(N-1)/2}\nonumber\\
&-&\frac{\Gamma}{a}\sum_{\ell=1}^{N-1} \left[\sin(a\ell)-i\cos(a\ell)\right]\frac{\cos(x\ell)}{\ell}A_{N-\ell}(x,t)\label{eq:A2}
\end{eqnarray}
For an infinite chain,
\[
\lim_{N\rightarrow\infty}\frac{\sin[(x-a)N/2]}{\sin[(x-a)/2]}=2\pi\delta(x-a)
\]
and
\begin{eqnarray}
\frac{\partial A_\infty(x,t)}{\partial t}&=& \left[i\Delta_0 +i\frac{\Omega_{\infty}(x)}{2}-\frac{\Gamma_{\infty}(x)}{2}\right]A_\infty(x,t)-i\pi\Omega_0\delta(x-a)\label{eq:A:infinite}
\end{eqnarray}
where $\Gamma_\infty$ and $\Omega_\infty$ are defined in Eqs.(\ref{approx:scalar:BZ}) and (\ref{WW}).
The solution of Eq.(\ref{eq:A:infinite}) is
\begin{eqnarray}
A_\infty(x,t)&=&A_\infty(x,0)e^{(i\Delta_0 +i\Omega_{\infty}(x)/2-\Gamma_{\infty}(x)/2)t}\nonumber\\
&+&\frac{2\pi\Omega_0\delta(x-a)}{2\Delta_0+\Omega_{\infty}+i\Gamma_{\infty}}\left(1-e^{(i\Delta_0 +i\Omega_{\infty}(x)/2-\Gamma_{\infty}(x)/2)t}\right).\label{A:infinite}
\end{eqnarray}

\section{Generation of subradiance}

Based on the previous expressions, we now discuss how subradiance can be generated studying the temporal evolution from some initial conditions of the probability amplitudes $\beta_j$. The case of the excitation by an incident laser field will be discussed in subsection \ref{laser}.

\subsection{Single excited atom}\label{single}

\begin{figure}
      \centerline{\scalebox{0.4}{\includegraphics{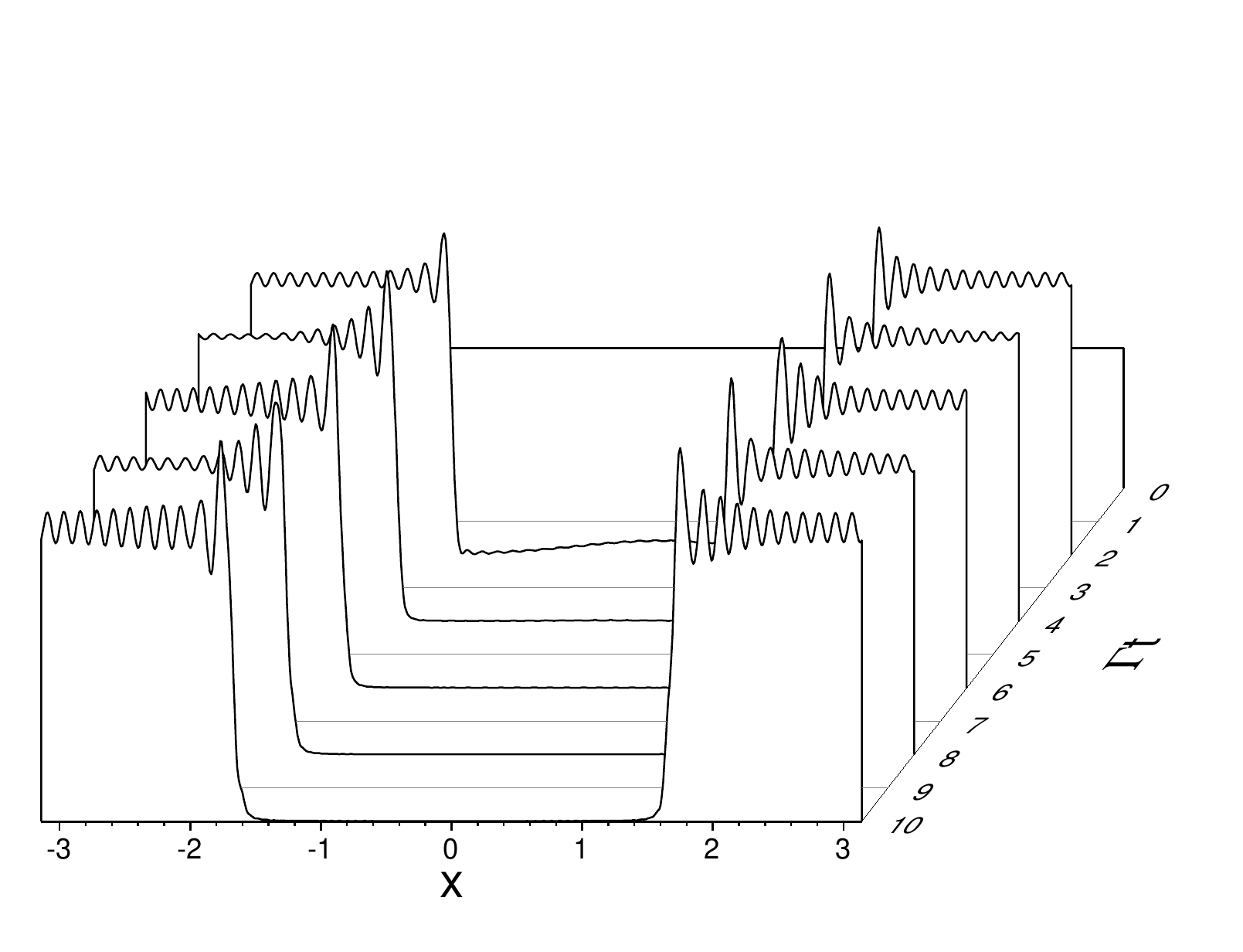}}}
        \caption{$P(x,t)$ vs $x$ for $\Gamma t=0,\dots,10$,  $a=\pi/2$ and $N=100$, with initially a single atom excited in the middle of the chain, obtained from the vectorial model with $\delta=\pi/2$.}\label{fig3}
\end{figure}

\begin{figure}
      \centerline{\scalebox{0.4}{\includegraphics{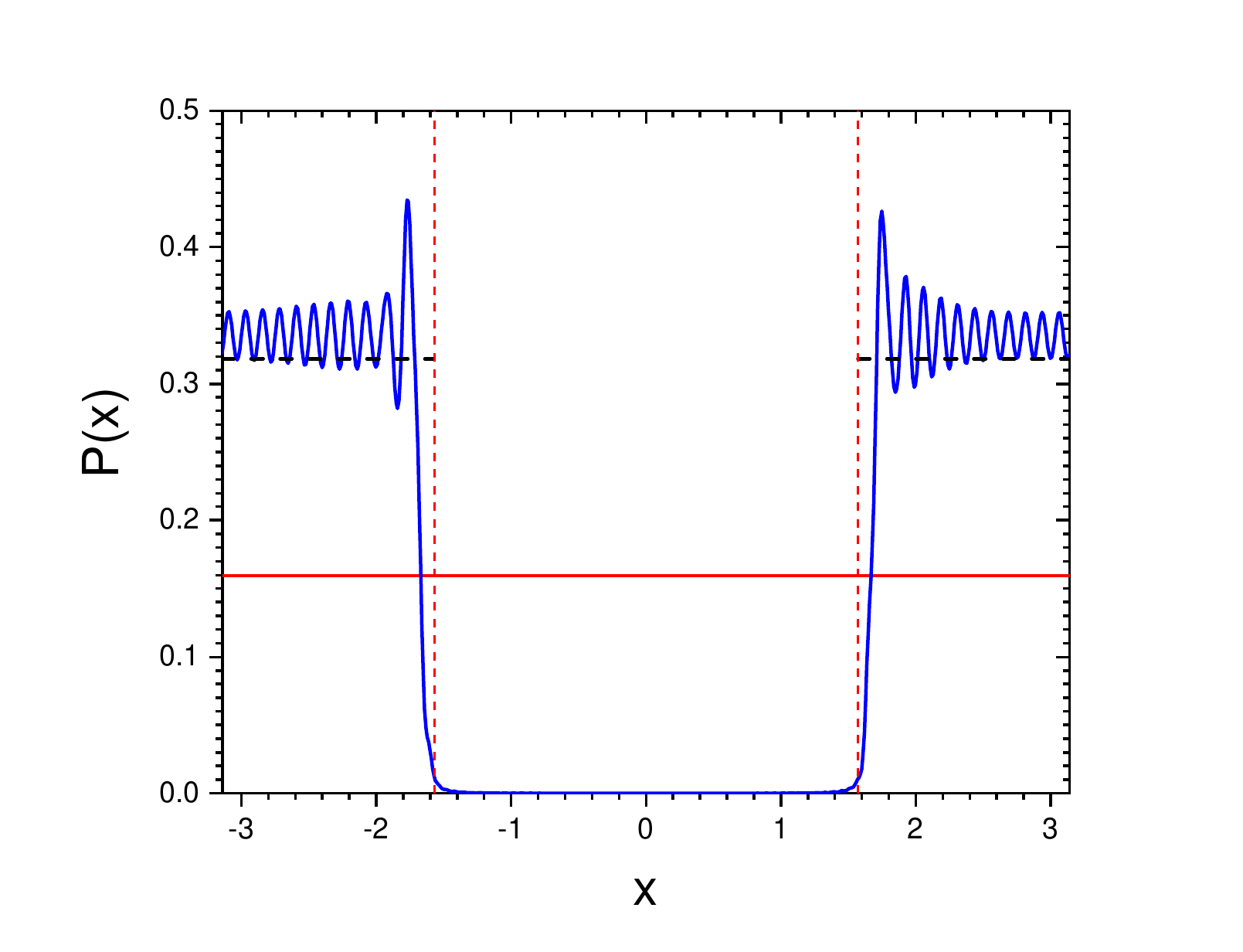}}}
        \caption{$P(x,t)$ vs $x$ (blue line) for $\Gamma t=10$ and the same parameters as in Fig.\ref{fig3}, with initially a single atom excited in the middle of the chain. The dashed line is the analytical result $P(x,\infty)=1/[2(\pi-a)]$ for an infinite chain, while the red line is the initial value $P(x,0)=1/2\pi$.}
        \label{fig4}
\end{figure}

\begin{figure}
      \centerline{\scalebox{0.4}{\includegraphics{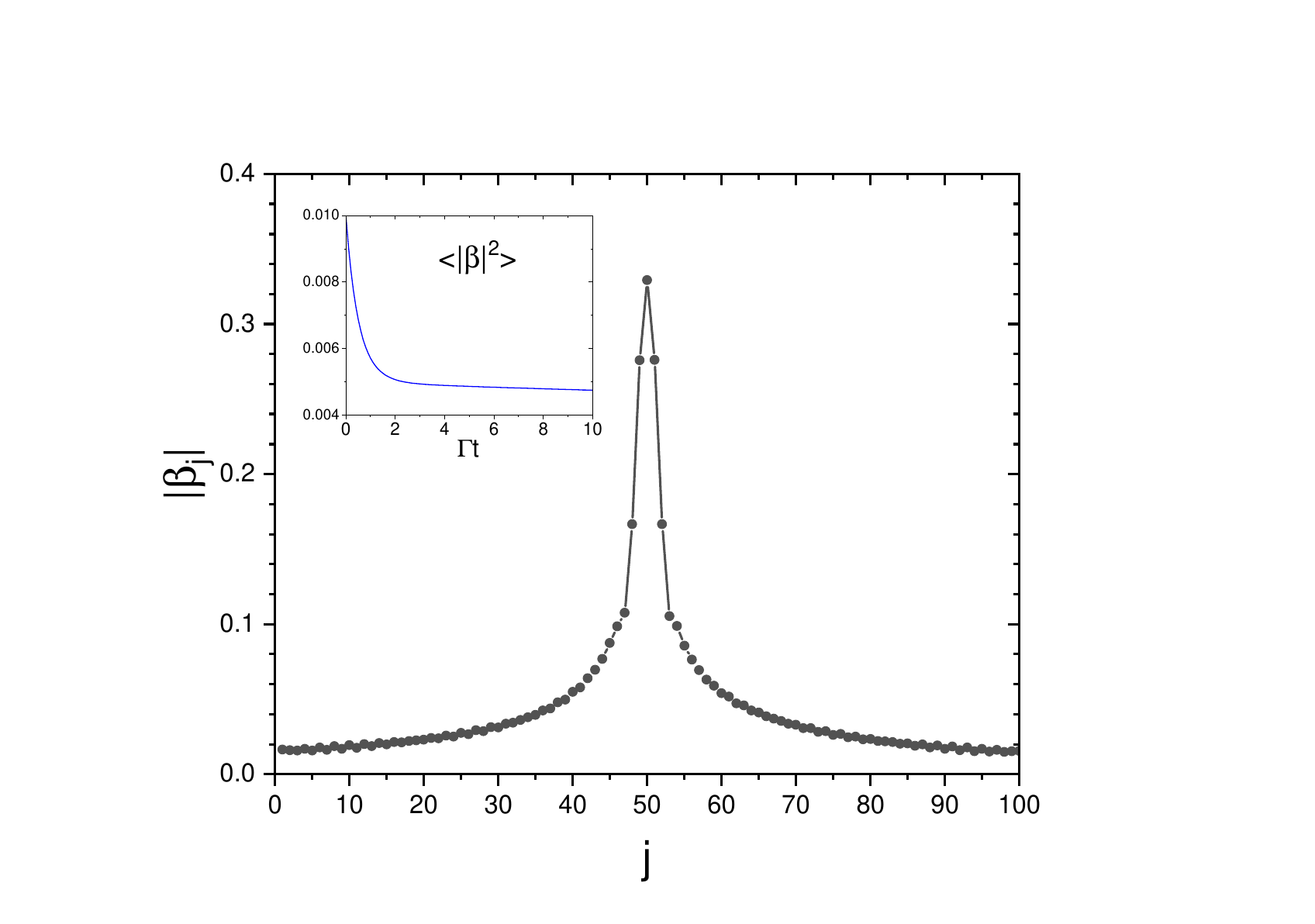}}}
        \caption{$|\beta_j|$ vs $j$ for $N=100$ and $a=\pi/2$, at $\Gamma t=10$, for the case of Fig.\ref{fig4}. The inset shows the average excitation probability $\langle|\beta|^2\rangle$ vs time.}
        \label{fig5}
\end{figure}

As a first example, we consider a chain of $N=100$ atoms with  $a=\pi/2$, no driving laser, $\Omega_0=0$ and $\Delta_0=0$, and a single initial atom excited in the middle of the chain, with $\beta_{N/2}=1$ and all the others $\beta_j$ equal to zero. From Eq.(\ref{Pk}), the initial probability to be in the state $|x\rangle$ is $P(x,0)=1/(2\pi)$ (where $x=kd$) i.e. it is uniform.  Fig.\ref{fig3} shows $P(x,t)$ vs $x$ at different times, from $t=0$ until $t=10/\Gamma$, obtained solving Eq.(\ref{beta_dot:vec}) for $\delta=\pi/2$: after few time units, the probability becomes zero for $|x|<a$ and different from zero in the subradiant interval $a<|x|<\pi$.
For an infinite chain, $|A_{\infty}(x,t)|^2=\exp[-\Gamma_{\infty}(x)t]$ where $\Gamma_{\infty}(x)=0$ for $a<|x|<\pi$.
In the limit $t\rightarrow\infty$, $P(x,t)\rightarrow 0$ for $|x|<a$ and $P(x,t)\rightarrow \frac{1}{2(\pi-a)}$ for $a<|x|<\pi$. Fig.\ref{fig4} shows $P(x,t)$ at $t=10/\Gamma$ and $a=\pi/2$ (blue continuous line), obtained for the same parameters as in Fig.\ref{fig3}, together with the value $P(x,\infty)=1/[2(\pi-a)]$ obtained for infinite chain (dashed line).
Hence, a single excited atom generates a subradiance state with a probability $P(x)$ which for an infinite chain is uniform in the subradiance spectral region $a<|x|<\pi$. It is interesting to see the distribution of the dipole amplitudes for the case of Fig.\ref{fig4}: Fig.\ref{fig5} shows $|\beta_j|$ vs $j$ at $t=10/\Gamma$, where the inset shows the average excitation probability $\langle|\beta|^2\rangle$ vs time. The initial excitation for the atoms with $j=N/2$ spread among the adjacent atoms, to a final distribution generating subradiance. Also if not visible in the Fig.\ref{fig5}, it is expected that the destructive interference among the atoms inhibit the spontaneous decay of the excitation, as it will be discussed in the next section.

\subsection{The most subradiant state}

\begin{figure}
      \centerline{\scalebox{0.6}{\includegraphics{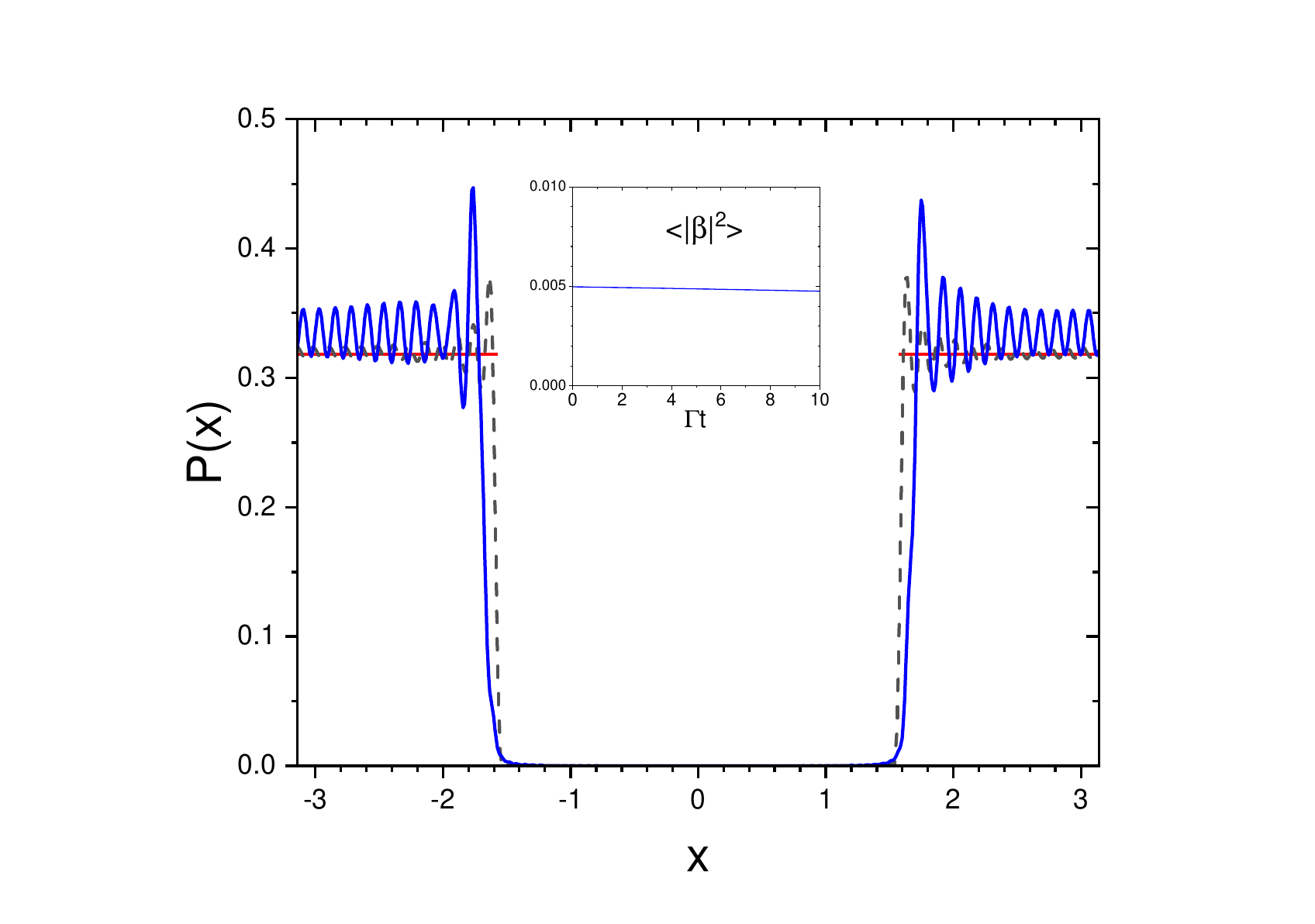}}}
        \caption{Probability density distribution $P(x,t)$ at $t=0$ (black-dashed line) and at $\Gamma t=10$ (blue line), for $N=100$, $a=\pi/2$ and $\delta=\pi/2$, obtained solving numerically Eq.(\ref{beta_dot}) with the initial condition (\ref{beta:sub}). The red line is the case of an infinite chain.  The inset shows that the average excitation probability $\langle|\beta|^2\rangle$ is almost constant.}
        \label{fig7}
\end{figure}

Since an initial uniform probability $P(x)$ generates asymptotically a subradiant distribution which is uniform for an infinite chain, as observed in the previous case, we are now interested to obtain the values of $\beta_j$ which are generating such subradiant distribution. We assume
\begin{equation}\label{A:sub}
    A_N(x)=e^{-ix(N/2-1)}
   	\left\{
    \begin{array}{lll}
    1 & \mbox{if}~a<&|x|<\pi\\
    0 & \mbox{if}~ &|x|<a.
\end{array}
	\right.
\end{equation}
which describes a subradiant state, with zero probability distribution in the superradiant region $|x|<a$ and uniform distribution in the subradiant region $a<|x|<\pi$. Calculating the single atom probability amplitude $\beta_j$ we obtain
\begin{equation}
  \beta_j=\frac{1}{2\pi}\int_{-\pi}^{\pi} e^{ix(j-1)}A_N(x) dx=\left\{
    \begin{array}{lll}
    1-\frac{a}{\pi} & \mbox{if}~&j=\frac{N}{2}\\
    &\\
    -\frac{\sin[a(j-N/2)]}{\pi(j-N/2)} & \mbox{if}~&j\neq \frac{N}{2}
\end{array}
	\right.\label{beta:sub}
  \end{equation}  
The $\beta_j$ as defined in Eq.(\ref{beta:sub}) reproduce the distribution amplitude (\ref{A:sub}) only in the limit $N\rightarrow\infty$ (see Appendix \ref{App:2}), since the states $|k\rangle$ are not orthogonal. 
Fig.\ref{fig7} shows the probability density distribution $P(x,t)$ at $t=0$ (black-dashed line) and at $\Gamma t=10$ (blue continuous line), obtained solving Eq.(\ref{beta_dot:vec}) with the initial condition (\ref{beta:sub}), $N=100$, $a=\pi/2$ and $\delta=\pi/2$. For an infinite chain, $P(x)=1/[2(\pi-a)]$ for $a<|x|<\pi$ and zero for $|x|<a$, as obtained from Eq.(\ref{A:sub}) (red line in Fig.\ref{fig7}). Notice the similarities between Fig.\ref{fig4} for the single initially excited atom and Fig.\ref{fig7} for the initial state (\ref{beta:sub}). We conclude that the state described by Eq.(\ref{beta:sub}) represents the 'most subradiant' state for a finite chain of $N$ atoms, with a purely subradiant spectrum in the limit of an infinite chain. 

\subsection{Atoms driven by a laser.}\label{laser}

\begin{figure}
      \centerline{\scalebox{0.4}{\includegraphics{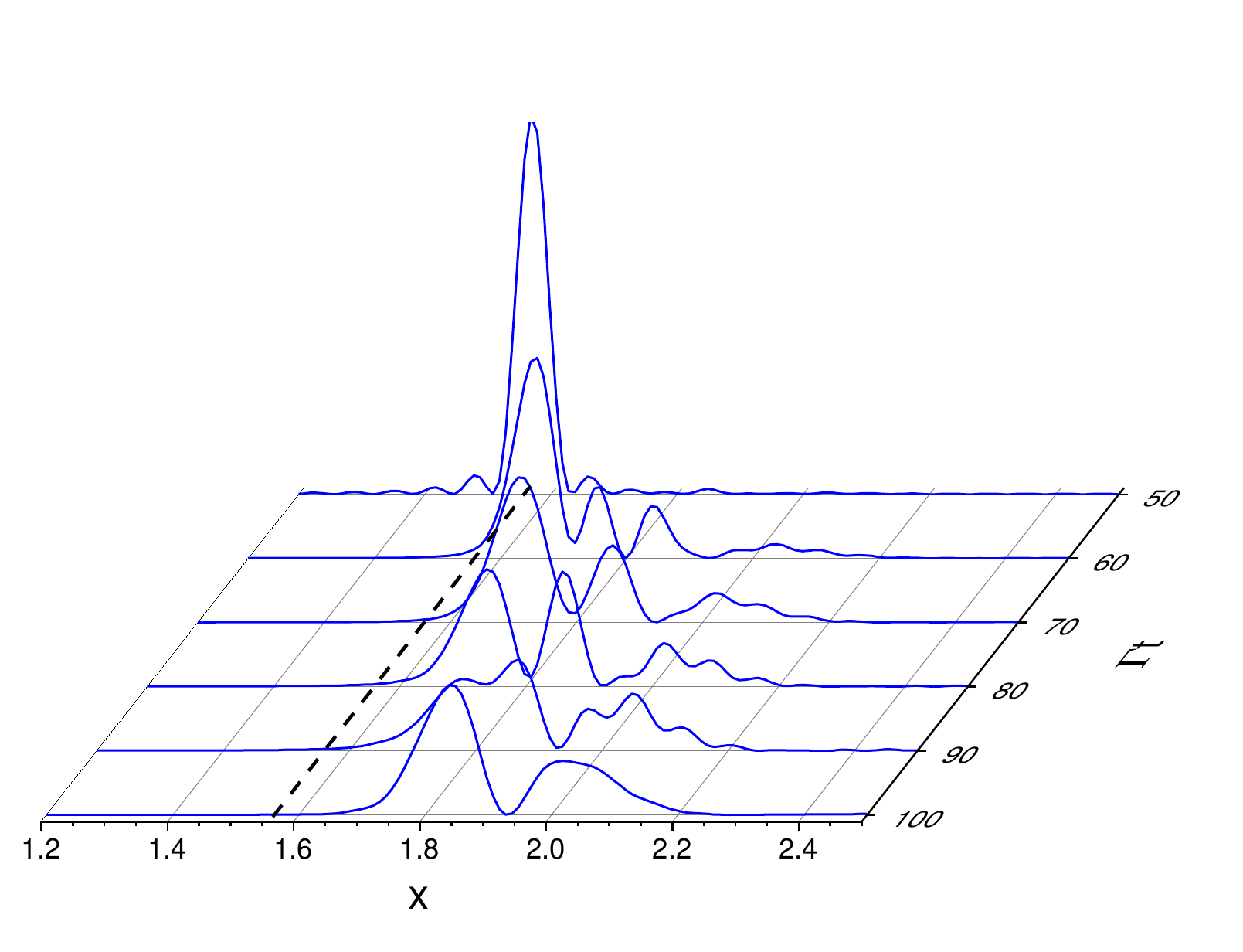}}}
        \caption{$P(x,t)$ vs $x$ at different times after the laser has been switched off, for a chain of $N=100$, $a=\pi/2$ and $\delta=\pi/2$, driven by detuned laser  with $\Omega_0=0.1\Gamma$ and $\Delta_0=10\Gamma$. The dashed line is the value $x=a$. The laser is switched off at $\Gamma t=50$.}
        \label{fig8}
\end{figure}

\begin{figure}
      \centerline{\scalebox{0.4}{\includegraphics{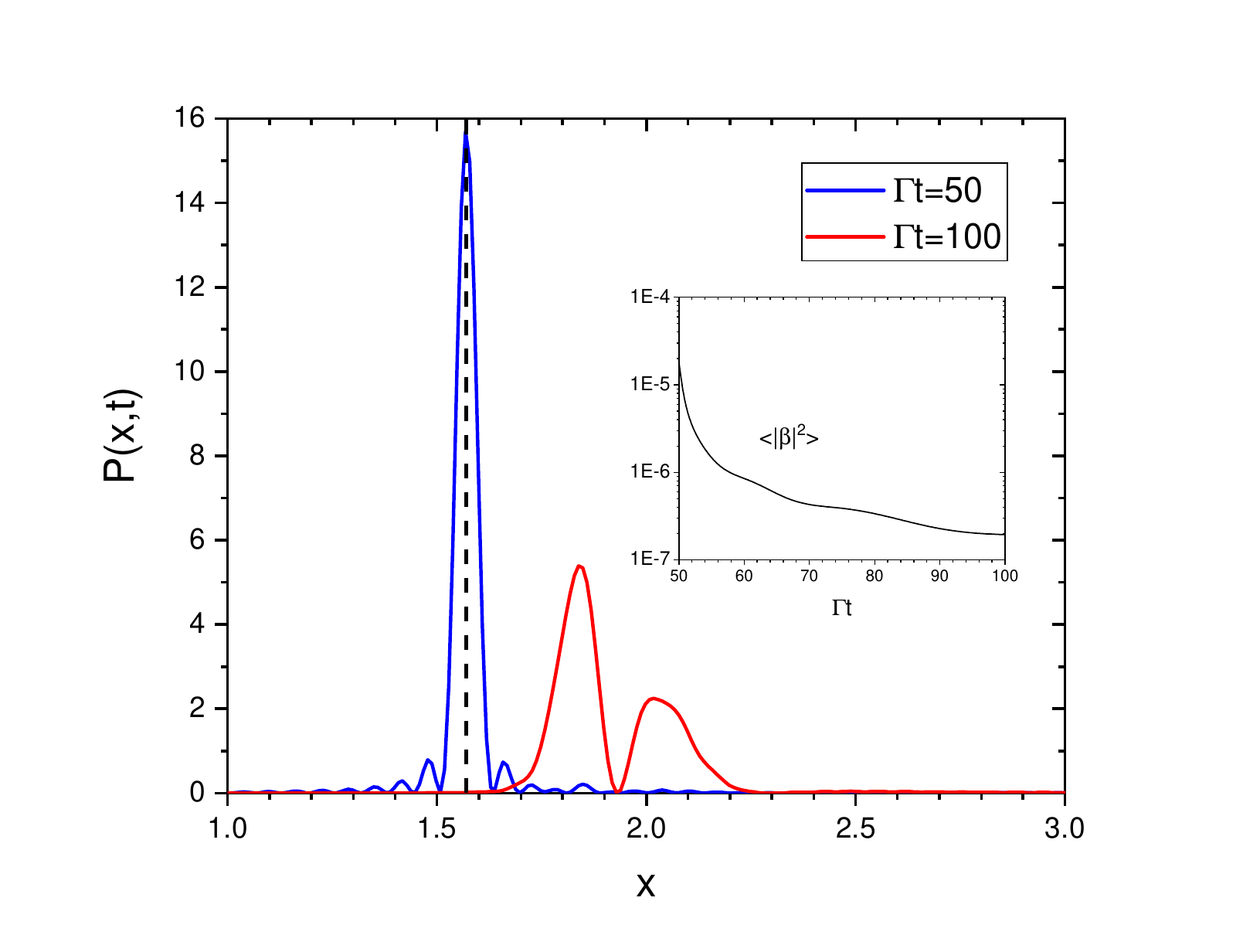}}}
        \caption{$P(x,t)$ vs $x$ at the switch-off time $\Gamma t=50$ (blue line) and at $\Gamma t=100$ (red line), for the same parameters as in Fig.\ref{fig8}. The vertical dashed line indicate the value $x=a=\pi/2$. The black dashed line is at $x=a$. The inset shows the average excitation probability $\langle|\beta|^2\rangle$ vs time.}
        \label{fig9}
\end{figure}

We now study the subradiance generation when the atoms are excited by an external laser field and then switched off.
As an example, we consider a chain of $N=100$ atoms with $a=\pi/2$ and $\delta=\pi/2$, weakly driven by a detuned laser, with $\Omega_0=0.1\Gamma$ and $\Delta_0=10\Gamma$. The atoms, initially unexcited (i.e. $\beta_j(0)=0$ for $j=1,\dots,N$) are driven by the laser up to $t_0=50/\Gamma$, after which the laser is switched off. Fig.\ref{fig8} shows $P(x,t)$ vs $x$ and at different times $t$ after the laser is switched off, obtained by solving the vectorial model of Eq.(\ref{beta_dot:vec}).
The distributions $P(x,t)$ at the laser switch-off time $t=50/\Gamma$ (blue continuous line) and at $t=100/\Gamma$ (red line), are shown in Fig.\ref{fig9}: we see that the driving laser on brings  the atoms close to a Timed-Dicke state, $|k_0\rangle$ \cite{Scully2006} (dashed line $x=a$ in Fig.\ref{fig9}), with a width inversely proportional to the chain's length $dN$. The inset of Fig.\ref{fig9} shows that, after the initial fast decay, subradiance manifests itself in a slow decay of the excitation. At first, the subradiant decay is not purely exponential, since several modes decay simultaneously. For longer times, it then ends up with a pure exponential decay when only one long-lived mode dominates. However, the precise evaluation of the decay rate can be problematic due to the general non-exponential decay. In our approach, we determine the precise distribution of the subradiant modes: as it can be observed from the red line of Fig.\ref{fig9}, at later times after the laser switch-off the distribution $P(x,t)$ is mostly in the subradiant region, $x>a$. The distribution is broad, so there is not a single subradiant mode dominating. As a consequence, the decay is not purely exponential.

In the case of infinite chain, Eq.(\ref{A:infinite}) gives
\begin{equation}
|A_{\infty}(x,\infty)|^2=\frac{2N(\pi\Omega_0)^2}{(2\Delta_0+\Omega_{\infty})^2+\Gamma_\infty^2}\delta(x-a),
\end{equation}
where we used the relation
\[
\lim_{N\rightarrow\infty}\frac{\sin^2[(x-a)N/2]}{\sin^2[(x-a)/2]}=\pi N\delta(x-a).
\]
Hence, for a driven infinite chain the asymptotic spectrum is $P(x,\infty)\propto\delta(x-a)$ and no subradiance occurs. To observe subradiance, we need a finite chain, as seen in Fig. \ref{fig8} and \ref{fig9}: in the detuned case, the finite width  of the driving term  (second term in r.h.s. of Eq.(\ref{eq:A2}) and blue line in Fig.\ref{fig9}) is proportional to $1/N$ and is responsible for the subradiant components of the spectrum until the laser is on, which subsequently evolve  without reaching a steady-state value. From the above analysis, the possibility to have access to the full spectral distribution of the subradiant modes is clearly more advantageous than observing the time decaying excitation to obtain the subradiant decay rate, as done for instance in Ref.\cite{Bienaime2012,Guerin2016}.

\begin{figure}
      \centerline{\scalebox{0.4}{\includegraphics{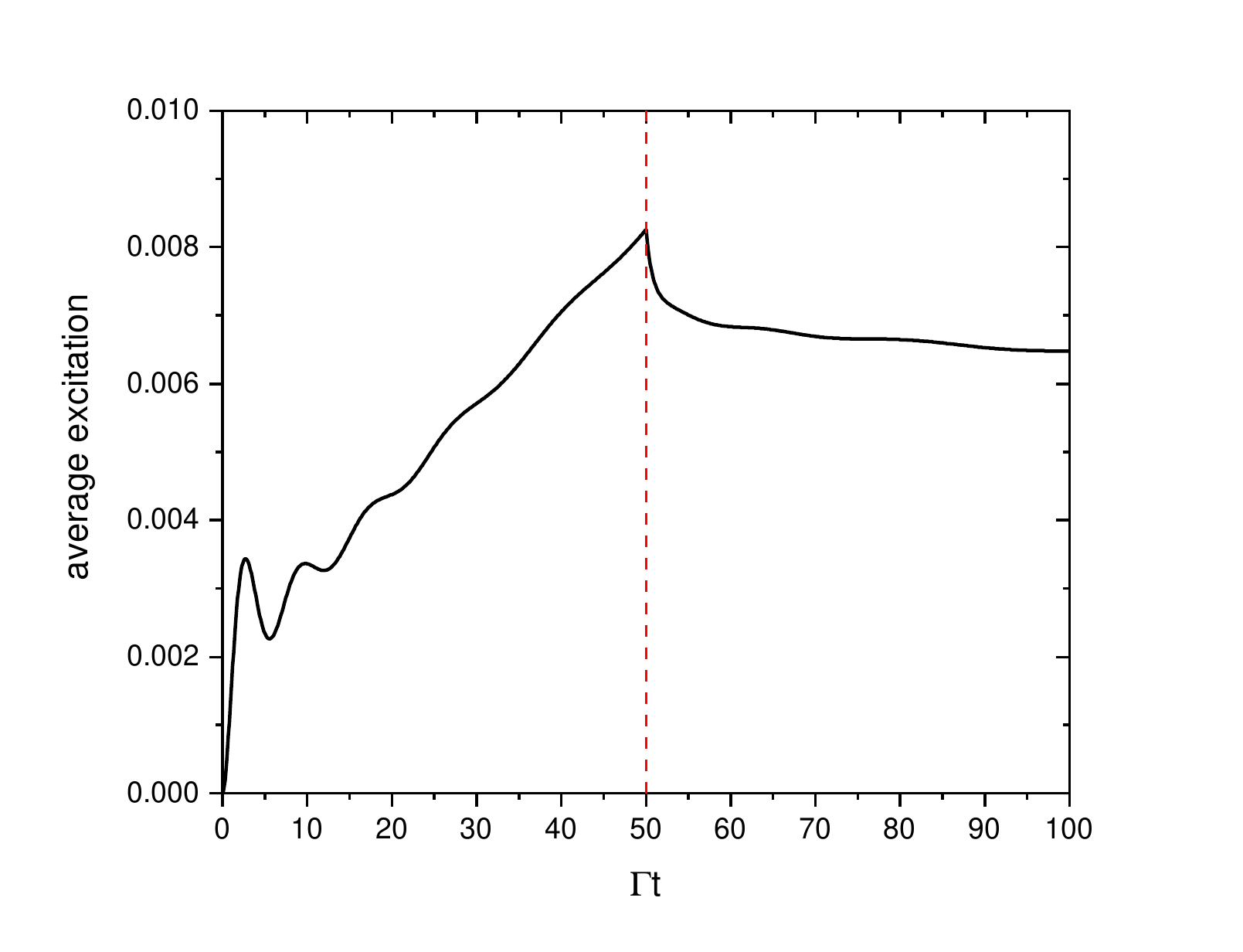}}}
        \caption{Average excitation, $\langle|\beta|^2\rangle$, vs $\Gamma t$ of a chain of $N=100$ with $a=\pi/2$, driven by a continuous resonant laser field, with $\Omega_0=0.1$ and $\Delta_0=0$, switched off at $\Gamma t=50$ (vertical dashed line).}
        \label{fig10}
\end{figure}

\begin{figure}
    \centerline{\scalebox{0.4}{\includegraphics{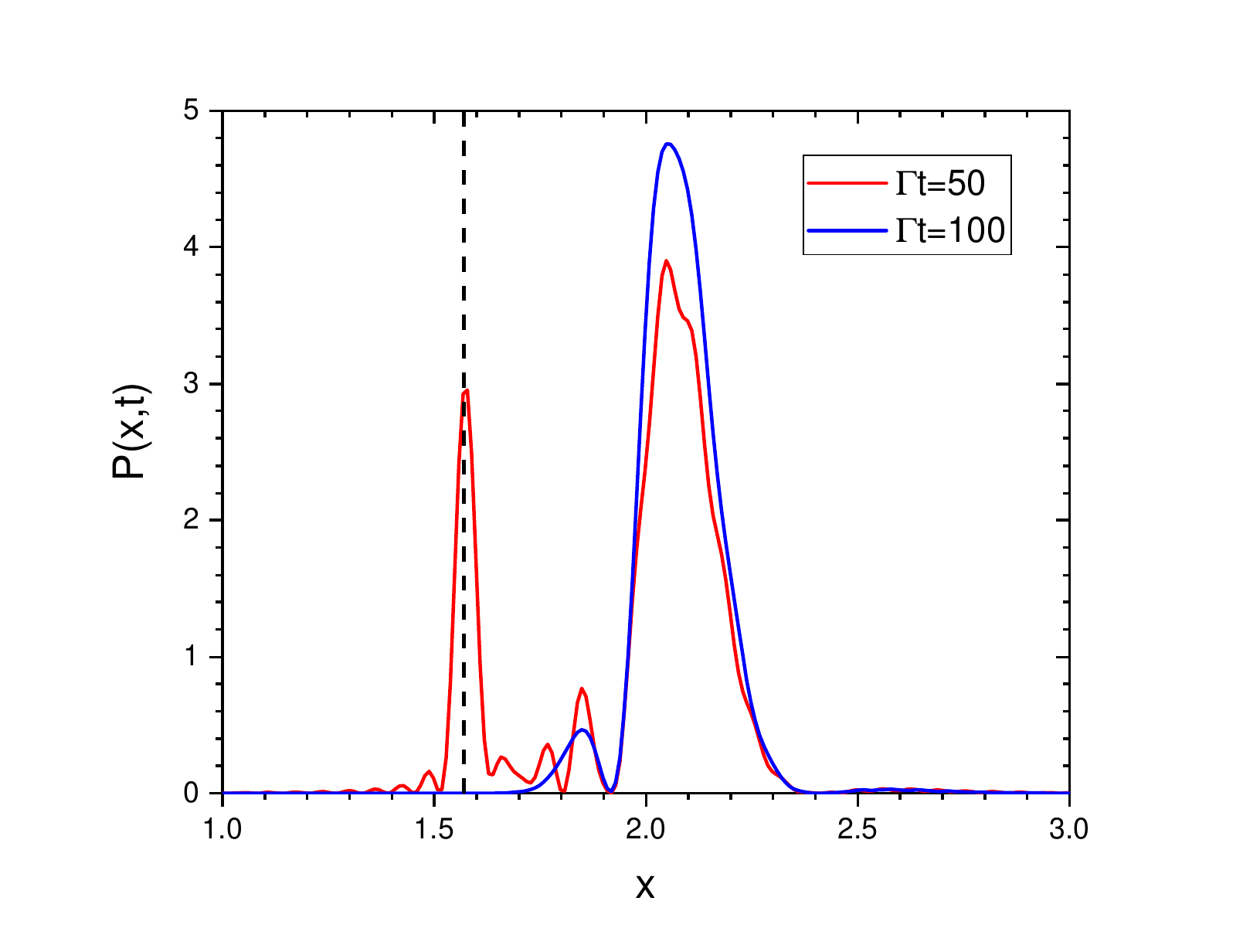}}}
        \caption{$P(x,t)$ vs $x$ at the laser switch-off time, $t=50/\Gamma$ (red line) and at $t=100/\Gamma$ (blue line), for $a=\pi/2$, $N=100$, $\Omega_0=0.1$ and $\Delta_0=0$. The dashed line is the value $x=a$.}
        \label{fig11}
\end{figure}

We now consider again the same chain of $N=100$ atoms with $a=\pi/2$, but driven by a resonant laser, with $\Omega_0=0.1$ and $\Delta_0=0$. The most interesting case is that of the scalar model, obtained assuming $\delta=54.37^\circ$ in the vectorial equations (\ref{beta_dot:vec}):  Fig.\ref{fig10} shows the average excitation, $\langle|\beta|^2\rangle$ vs $\Gamma t$, where the drive field is switched off at $\Gamma t=50$ (vertical dashed line). The average excitation grows almost linearly when the laser is turned on, typical for a diffusive regime \cite{Holstein1947,Labeyrie2003}. After the laser is off, the excitation decays very slowly, showing that the excitation remains trapped in the atomic chain. We can understand this peculiar behavior by observing the probability density $P(x,t)$ in Fig\ref{fig11} at the laser switch-off time, $t=50/\Gamma$ (red line) and at $t=100/\Gamma$ (blue line). Contrarily to the detuned case, at resonance the subradiant region of the spectrum is already populated when the laser is on (red line of  Fig.\ref{fig11}). After the laser has been switched off, at $\Gamma t=100$, $P(x,t)$ it remained almost the same, with only the radiating components, for $x<a$, decayed (blue line of  Fig.\ref{fig11}). Since now the spectrum is completely in the subradiant region $x>a$, the decay rate is almost zero and the atoms remain excited for a sufficiently long time.

\section{Radiated Intensity}

\begin{figure}
    \centerline{\scalebox{1.2}{\includegraphics{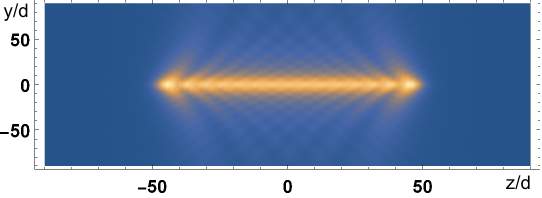}}}
        \caption{Field intensity (arb. units) in the plane $(y/d,z/d)$ at $x=5d$ emitted by a chain of $N=50$ atoms, with $kd=1$, along the $z$-axis, centered at $x=y=0$ and uniformly excited, $\beta_j=1/\sqrt{N}$. We observe that the field is radiated transversely to the chain.}
        \label{fig12}
\end{figure}

\begin{figure}
    \centerline{\scalebox{1.2}{\includegraphics{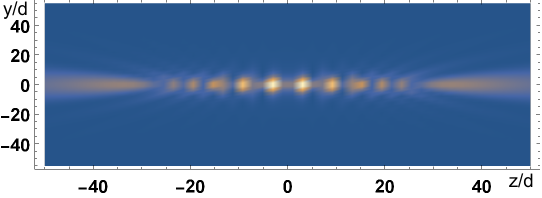}}}
        \caption{Same as in Fig.\ref{fig13}, but for the $\beta_j$ given by Eq.(\ref{beta:sub}), referred as the 'most subradiant state'. We observe that the field is largely evanescent transverse to the chain, while most of the energy is radiated out at the ends of the chain.}
        \label{fig13}
\end{figure}

The following important question arises: may the probability $P(x,t)$  be determined measuring the scattered intensity at a certain angle $\theta$ with respect to the chain's axis? We know that the scattered field appears as a sum of wavelets radiated by the atomic dipoles, with polarization component of the electric field
\begin{equation}
        E_\alpha(\mathbf{r},t)=\frac{\hbar}{ie\mu}\sum_{\beta}\sum_{j=1}^N G_{\alpha,\beta}(k_0|\mathbf{r}-\mathbf{r}_j|)\beta_j(t)
        \label{Es}
\end{equation}
where $G_{\alpha,\beta}(\mathbf{r})$ is defined by Eq.(\ref{Gaa}). 
In the far-field limit, one has $k_0|\mathbf{r}-\mathbf{r}_j|\approx k_0r-\mathbf{k}\cdot \mathbf{r}_j$, where $\mathbf{k}=k_0(\sin\theta\cos\phi,\sin\theta\sin\phi,\cos\theta)$ and $\mathbf{r}_j=d(j-1)\hat{\mathbf{e}}_z$, so the field of Eq.(\ref{Es}) radiated in a direction $\mathbf{\hat n}$
reads
\begin{equation}
        \mathbf{E}(\theta,t)\approx\frac{3\hbar\Gamma}{2e\mu}\mathbf{\hat n}\times(\mathbf{\hat n}\times \mathbf{\hat e})\frac{e^{ik_0r}}{k_0r}\sum_{j=1}^N e^{-ik_0d\cos\theta(j-1)}\beta_j(t)
        \label{Es:far}
\end{equation}
where $\mathbf{\hat e}$ is is the unit polarization vector of the dipoles, and the scattered intensity is
\begin{equation}
         I_s(\theta,t)\propto\left|\sum_{j=1}^N e^{-ik_0d\cos\theta(j-1)}\beta_j(t)\right|^2\propto P(k_z,t)
         \end{equation}
where $k_z=k_0\cos\theta$. Hence, the atoms radiate out of the chain's axis for $|k_z|<k_0$. The subradiant region $|k_z|>k_0$  is not accessible by the scattered field, since it would be $\cos\theta>1$ and the electromagnetic field is evanescent in the directions transverse to the chain, since $k_\perp=ik_0\sqrt{\cos^2\theta-1}=i\xi$. Very few photons are emitted outside the chain's axis direction (anyone in the case of an infinite chain).
However, from the radiated intensity it is possible to see if the atomic state is subradiant or not, observing if the atoms are emitting in a direction out of the axis' chain. This can be seen in Figures \ref{fig12} and \ref{fig13}, where we plot the field intensity $I_s\propto |E_\alpha|^2$ (where $E_\alpha$ is determined by Eq.(\ref{Es})) in the plane $x=5d$, emitted by a chain of $N=50$ atoms with $kd=1$, centered at $x=y=0$, for two different atomic distributions. Fig. \ref{fig12} shows a case of uniform excitation, with $\beta_j=1/\sqrt{N}$ such that $P(k_z)\propto N{\rm sinc}^2(k_zdN/2)$ and the probability distribution is peaked around $k_z=0$: in this case the atoms emit out of the chain's axis. 
Fig. \ref{fig13} shows the emission from the most subradiant state, with $\beta_j$ given by Eq.(\ref{beta:sub}), such that $P(k_z)\approx 0$ for $k_z<k_0$: the field is evanescent and does not propagate out of the chain, remaining confined along the chain; since the chain is finite, most of the energy is radiated out at the ends of the chain \cite{Asenjo2017}.

\section{Conclusions}

In conclusion, we have discussed analytically and numerically how subradiance can emerge from the evolution of the dynamics of $N$ two-level atoms in the single-excitation configuration along a linear chain. In the first part, we have characterized the spectrum of the decay rates and frequency shifts of the system, identifying the regions of the spectrum where spontaneous emission is enhanced or inhibited, up to a complete suppression in the case of an infinite chain. 
We proceeded first by obtaining a relation between the spectrum of emission and the single-particle amplitudes, whose evolution can be determined by solving the coupled-dipole equations. Then we have studied how different initial excitations evolve toward a subradiant state. A single-excited atom leads to an almost uniform population of subradiant modes. This has suggested the idea that the atomic configuration leading to this uniform population can be calculated directly, obtaining what we named the 'most subradiant state'. Then, we investigated how subradiance may be induced by a driving laser, which excite the atoms and successively is switched off, such that the long-lived subradiant modes survive for long times. Finally, we found the relation between spontaneous emitted intensity and subradiance. Subradiance is characterized by a suppression of the emission in the direction transverse to the chain axis. This analysis may be useful to envisage strategies to detect subradiance in ordered systems by measuring the radiation out of the lattice. The results obtained here for a linear chain can be extended to 2D a 3D lattices.

\appendix

\section{Proof of Eq.(\ref{Sk})}\label{eq:4}
To prove Eq.(\ref{Sk}), we write
\begin{eqnarray}
\Gamma_{jm}&=&\frac{\Gamma}{2}\left\langle e^{-i\mathbf{k}\cdot (\mathbf{r}_j-\mathbf{r}_m)} 
+\mathrm{c.c.}\right\rangle_\Omega\nonumber\\
&=&\frac{\Gamma}{8\pi}\int_0^{2\pi}d\phi\int_0^\pi \sin\theta\left[e^{-ik_0r_{jm}\cos\theta}+\mathrm{c.c.}\right]d\theta\nonumber\\
&=&\frac{\Gamma}{2}\int_0^\pi \sin\theta\cos(k_0r_{jm}\cos\theta)d\theta=\frac{\sin(k_0r_{jm})}{k_0r_{jm}}
\end{eqnarray}
where $r_{jm}=|\mathbf{r}_j-\mathbf{r}_m|$.

\section{Equation for $A_N(x,t)$}\label{eq:A}

The equation for the temporal evolution of the probability amplitude $A_N(x,t)$ (where $x=kd$) can be obtained from Eqs.(\ref{beta_dot}) and (\ref{def:A}):
\begin{eqnarray}
\frac{\partial A_N(x,t)}{\partial t}&=& \left(i\Delta_0-\frac{\Gamma}{2}\right) A_N(x,t)
-i\frac{\Omega_0}{2}\frac{\sin[(x-a)N/2]}{\sin[(x-a)/2]}e^{-i(x-a)(N-1)/2}\nonumber\\
&-&\frac{\Gamma}{2}\sum_{j=1}^N\sum_{m=1\atop m\neq j}^N\left[
\frac{\sin a|j-m|}{a|j-m|}-i\frac{\cos a|j-m|}{a|j-m|}\right]
e^{-ix(j-m)}e^{-ix(m-1)}\beta_m(t)\nonumber\\
\end{eqnarray}
where $x=kd$ and $a=k_0d$.
The third term can be written, introducing the index $\ell=j-m$, as
\begin{eqnarray}
&&\sum_{\ell=-(N-1)\atop\ell\ne 0}^{N-1}\left[\frac{\sin a|\ell|}{a|\ell|}-i\frac{\cos a|\ell|}{a|\ell|}\right]e^{-ix\ell}\sum_{m=1}^{N-|\ell|}e^{-ix(m-1)}\beta_m(t)\nonumber\\
&=&\frac{2}{a}\sum_{\ell=1}^{N-1}[\sin(a\ell)-i\cos(a\ell)]\frac{\cos(x\ell)}{\ell}A_{N-\ell}(x,t),
\end{eqnarray}
so that
\begin{eqnarray}
\frac{\partial A_N(x,t)}{\partial t}&=& \left(i\Delta_0-\frac{\Gamma}{2}\right)A_N(x,t)-i\frac{\Omega_0}{2}\frac{\sin[(x-a)N/2]}{\sin[(x-a)/2]}e^{-i(x-a)(N-1)/2}\nonumber\\
&-&\frac{\Gamma}{a}\sum_{\ell=1}^{N-1} \left[\sin(a\ell)-i\cos(a\ell)\right]\frac{\cos(x\ell)}{\ell}A_{N-\ell}(x,t)
\end{eqnarray}

\section{Probability amplitude for the subradiant state}\label{App:2}
Assuming the subradiant state with
\begin{equation}
    \beta_j=
   	\left\{
    \begin{array}{ll}
    1-\frac{a}{\pi} & \mbox{if}~j=\frac{N}{2}\\
    &\\
    -\frac{\sin[a(j-N/2)]}{\pi(j-N/2)} & \mbox{if}~j\neq \frac{N}{2}
\end{array}
	\right.
\end{equation}
we can calculate the probability amplitude as
\begin{eqnarray}
A_N(x)&=&\sum_{j=1}^N e^{-ix(j-1)}\beta_j=e^{-ix(N/2-1)}\frac{\pi-a}{\pi}\nonumber\\
&-&\frac{1}{\pi}\sum_{j=1}^{N/2-1} \frac{\sin[a(j-N/2)]}{j-N/2}e^{-ix(j-1)}\nonumber\\
&-&\frac{1}{\pi}\sum_{j=N/2+1}^{N} \frac{\sin[a(j-N/2)]}{j-N/2}e^{-ix(j-1)}\nonumber\\
&=& \frac{e^{-ix(N/2-1)}}{\pi}\left\{
\pi-a-\sum_{m=1}^{N/2-1} \frac{\sin(am)}{m}e^{ixm}-\sum_{m=1}^{N/2} \frac{\sin
(am)}{m}e^{-ixm}
\right\}
 \end{eqnarray}
In the limit $N\rightarrow\infty$, apart for the phase global phase factor
\begin{eqnarray}
|A_\infty(x)|&=&\frac{1}{\pi}
\left|\pi-a-\sum_{m=1}^{\infty} \frac{\sin(am)}{m}\left(e^{ixm}+e^{-ixm}\right)\right|\nonumber\\
&=&\frac{1}{\pi}\left|\pi-a+\theta_2-\theta_1\right|
 \end{eqnarray}
 where
\begin{equation}
 \theta_{1,2}=\arctan\{\sin(x\pm a)/[1-\cos(x\pm a)]\}=\arctan[\cot[(x\pm a)/2]].
 \end{equation} 
 Since $\arctan[\cot(z)]=\pi/2-(z-m\pi)$ for $m\pi<z<(m+1)\pi$, then $\theta_2-\theta_1=a$ for $|x|>a$ and 
 $\theta_2-\theta_1=-(\pi-a)$ for $|x|<a$. Finally
 \begin{equation}
    |A_\infty(x)|=
   	\left\{
    \begin{array}{lll}
    1 & \mbox{if}~ &|x|>a\\
    0 & \mbox{if}~ &|x|<a.
\end{array}
	\right.
\end{equation}
Hence, we obtain the 'full subradiant state' (\ref{A:sub}) only in the limit of an infinite chain.


\begin{thebibliography}{999}
\bibitem[1]{Dicke1954} Dicke R. H., Coherence in spontaneous radiation processes.
{\em Phys. Rev.} {\bf 1954}, {\em 93}, 99.
\bibitem[2]{Lehmberg1970} Lehmberg R. H., Radiation from an N-Atom system, I. General
formalism. {\em Phys. Rev. A} {\bf 1970}, {\em 2}, 883.
\bibitem[3]{Bonifacio1971} Bonifacio R.; Schwendimann P.; Haake F. Quantum Statistical Theory of Superradiance I. {\em Phys. Rev. A} {\bf 1971}, {\em 4}, 302.
\bibitem[4] {Gross1982} Gross M.; Haroche S. Superradiance: An Essay on the
Theory of Collective Spontaneous Emission. {\em Phys. Rep.} {\bf 1982}, {\em 93}, 301.
\bibitem[5]{Bienaime2012} T. Bienaim\'{e}, N. Piovella, and R. Kaiser, Controlled Dicke subradiance from a large cloud of two-level systems.
{\em Phys. Rev. Lett.} {\bf 2012}, {\em 108}, 123602.
\bibitem[6] {Guerin2016} W. Guerin, M. O. Ara\`{u}jo, and R. Kaiser, Subradiance in a large cloud of cold atoms. {\em Phys. Rev. Lett.} {\bf 2016}, {\em 116}, 083601.
\bibitem[7]{Das2020} D. Das, B. Lemberger, and D. D. Yavuz, Subradiance and Superradiance-to-Subradiance Transition in Dilute Atomic Clouds. {\em Phys. Rev. A} {\bf 2020}, {\em 102}, 043708.
\bibitem[8]{Ferioli2021} G. Ferioli, A. Glicenstein, L. Henriet, I. Ferrier-Barbut, and A. Browaeys, Storage and release of subradiant excitations
in a dense atomic cloud. {\em Phys. Rev. X} {\bf 2021}, {\em 11}, 021031.
\bibitem[9]{Bettles2016} Bettles R.J.; Gardiner S.A.; Adams C.S. Cooperative eigenmodes and scattering in one-dimensional atomic arrays. {\em Phys. Rev. A} {\bf 2016}, {\em 94}, 043844.
\bibitem[10]{Facchinetti2016} Facchinetti G.; Jenkins S.D.; Ruostekoski J. Storing light with subradiant correlations in arrays of atoms. {\em Phys.Rev. Lett.} {\bf 2016}, {\em 117}, 243601.
\bibitem[11]{Asenjo2017} Asenjo-Garcia A.; Moreno-Cardoner M.; Albrecht A:; 
Kimble H. J.; Chang D. E. Exponential improvement in photon storage fidelities using subradiance and “selective radiance” in
atomic arrays. {\em Phys. Rev. X} {\bf 2017}, {\em 7}, 031024.
\bibitem[12]{Rui2020} Rui J.; Wei D.; Rubio-Abadal A.; Hollerith S.; Zeiher  J.; Stamper-Kurn D. M.; Gross C.; Bloch I.; A Subradiant Optical Mirror Formed by a Single Structured Atomic Layer. {\em Nature} {\bf 2020}, {\em 583}, 369.
\bibitem[13]{Cech2023} Cech M.; Lesanovsky I.; Olmos B. Dispersionless subradiant photon storage in one-dimensional emitter chains. {\em Phys. Rev. A} {\bf 2023}, {\em 108}, L051702.
\bibitem[14]{Piovella2024} Piovella N. Cooperative Decay of an Ensemble of Atoms in a One-Dimensional Chain with a Single Excitation.  {\em Atoms} {\bf 2024}, {\em 12}, 43.
\bibitem[15]{Bellando2014} Bellando L; Gero A.; Akkermans E.; Kaiser R. Cooperative effects and disorder: A scaling analysis of the spectrum of the effective atomic Hamiltonian. {\em Phys. Rev. A} {\bf 2014}, {\em 90}, 063822.
\bibitem[16]{Cottier2018} Cottier F.; Kaiser R,; Bachelard R. Role of disorder in super- and subradiance of cold atomic clouds. {\em Phys. Rev. A} {\bf 2018}, {\em 98}, 013622.
\bibitem[17]{Fofanov2021} Fofanov Y.A.; Sokolov I.M.; Kaiser R.; Guerin W. Subradiance in dilute ensembles: Role of pairs and multiple scattering. {\em Phys. Rev. A} {\bf 2021}, {\em 104}, 023705.
\bibitem[18] {Nienhuis1987} Nienhuis, G.; Schuller, F. Spontaneous emission and light scattering by atomic lattice models. {\em J. Phys. B: Atom. Mol. Phys.}  {\bf 1987}, {\em 20}, 23.
\bibitem[19]{Zoubi2010} Zoubi H.; Ritsch H. Metastability and Directional Emission Characteristics of Excitons in 1D Optical Lattices. 
{\em Europhys. Lett.}, {\bf 2010}, {\em 90}, 23001.
\bibitem[20]{Jenkins2012} Jenkins S.D.; Ruostekoski J. Controlled manipulation of light by cooperative response of atoms in an optical lattice. 
{\em Phys. Rev. A} {\bf 2012}, {\em 86}, 031602(R).
\bibitem[21]{Bettles2015} Bettles R. J. ; Gardiner S. A.; Adams C.S. Cooperative Ordering in Lattices of Interacting Two-Level Dipoles. {\em Phys. Rev. A } {\bf 2015}, {\em 92}, 063822.
\bibitem[22]{Needham2019} Needham J.A.; Lesanovsky I.; Olmos B. Subradiance-protected excitation transport. {\em New J. Phys.} {\bf 2019}, {\em 21}, 073061.
\bibitem[23] {Masson2020} Masson S.J.; Ferrier-Barbut I.; Orozco L.A.; Browaeys A.; Asenjo-Garcia A. Many-Body Signature of Collective Decay in Atomic Chains. {\em Phys. Rev. Lett.} {\bf 2020}, {\em 125}, 263601.
\bibitem[24]{Masson2022} Masson S.J.; Asenjo-Garcia A. Universality od Dicke superradiance in arrays of quantum emitters. {\em Nature Comm.} {\bf 2022}, {\em 13}, 2285.
\bibitem[25]{Ruostekoski2023} Ruostekoski J. Cooperative quantum-optical planar arrays of atoms. {\em Phys. Rev. A} {\bf 2023}, {\em 108}, 030101.
\bibitem[26]{Jenkins2017} Jenkins S.D.; Ruostekoski J.;, Papasimakis N:; Savo S.; Zheludev N.I. Many-Body Subradiant Excitations in Metamaterial Arrays: Experiment and Theory. {\em Phys. Rev. Lett.} {\bf 2017}, {\em 119}, 05390.
\bibitem[27]{Solano2017} Solano P.; Barberis-Blostein P.; Fatemi F.K.; Orozco L.A.; Rolston S.L. Super-radiance reveals infinite-range dipole interactions through a nanofiber. {\em Nat. Commun.} {\bf 2017}, {\em 8}, 1857.
\bibitem[28]{Jen2016} Jen H.H.; Chang M.-S.; Chen Y.-C. Cooperative single photon subradiant states. {\em Phys. Rev. A} {\bf 2016}, {\em 94}, 013803.
\bibitem[29]{Holzinger2020} Holzinger R.; Plankensteiner D.; Ostermann L.; Ritsch H. Nanoscale Coherent Light Source. {\em Phys. Rev. Lett.} {\bf 2020}, {\em 124}, 253603.
\bibitem[30]{Sonnefraud2010} Sonnefraud Y.; Verellen N.; H. Sobhani H.;,  Vandenbosch G.A.E.;
Moshchalkov V.V.; Dorpe P.V.; Nordlander P.; Maier S.A. Experimental realization of subradiant, superradiant, and Fano
resonances in ring/disk plasmonic nanocavities. {\em ACS Nano} {\bf 2010}, {\em 4}, 1664.
\bibitem[31]{McGuyer2015}  McGuyer B.H.; McDonald M.; Iwata G.Z.; Tarallo M.G.;
Skomorowski W.; Moszynski R.; Zelevinsky T. Precise study of asymptotic physics with subradiant ultracold molecules.
{\em Nat. Phys.} {\bf 2015}, {\em 11}, 32.
\bibitem[32]{Morsch2006} Morsch O.; Oberthaler M. Dynamics of Bose-Einstein condensates in optical lattices. {\em Rev. Mod. Phys.} {\bf 2006}, {\em 78}, 179.
\bibitem[33]{Hadzibabic2004} Hadzibabic Z.; Stock S.; Battelier B.; Bretin V.; Dalibard J. Interference of an Array of Independent Bose-Einstein Condensates. {\em Phys. Rev. Lett.} {\bf 2004}, {\em 93}, 180403.
\bibitem[34]{Akkermans2008} Akkermans E.; Gero A.; Kaiser R. Photon localization and Dicke superradiance in atomic gases.  {\em Phys. Rev. Lett.} {\bf 2008}, {\em 101}, 103602.
\bibitem[35]{Bienaime2013} Bienaim\'{e} T.; Bachelard R.; Piovella N.; Kaiser R. Cooperativity in light scattering by cold atoms. {\em Fortschritte der Physik} {\bf 2013}, {\em 61}, 377.
\bibitem[36]{Scully2006} Scully, M.O.; Fry, E.; Ooi, C.H.R.; Wodkiewicz, K. Directed Spontaneous Emission from an Extended Ensemble of N Atoms: Timing Is Everything. {\em Phys. Rev. Lett.} {\bf 2006}, {\em 96}, 010501.
\bibitem[37]{Scully2015} Scully M.O. Single photon subradiance: quantum control of spontaneous emission and ultrafast readout. {\em Phys. Rev. Lett.} {\bf 2015}, {\em 115}, 243602.
\bibitem[38]{Holstein1947} Holstein T. Imprisonment of Resonance Radiation in Gases. {\em Phys. Rev.} {\bf 1947}, {\em 72}, 1212.
\bibitem[39]{Labeyrie2003} Labeyrie G.; Vaujour E.; M\"{u}ller C.; Delande C.; Miniatura C.; Wilkowski D.; Kaiser R. Slow diffusion of light in a cold atomic cloud, {\em Phys. Rev. Lett.} {\bf 2003}, {\em 91}, 223904.
\end{thebibliography}
\end{document}